\documentclass[reprint,amsmath,amssymb,
 aps]{revtex4-2}
\usepackage{graphicx}
\graphicspath{{graphics/}}
\usepackage{epstopdf}  
\usepackage{ragged2e} 
\usepackage{indentfirst} 
\usepackage{setspace} 
\usepackage{dcolumn}
\usepackage{cases} 
\usepackage{bm}
\usepackage{latexsym} 
\usepackage{color} 
\usepackage[
bookmarks=true,         
linktocpage=true,       
hyperindex=true,        
colorlinks=true,        
linkcolor=blue, 
urlcolor=blue, 
anchorcolor=black, 
citecolor=blue]{hyperref} 
\usepackage{simplewick} 
\usepackage[compat=1.1.0]{tikz-feynman} 
\usepackage{slashed}
\makeatletter
\newcommand{\bal}{\@ifstar{\@bals}{\@bal}}
\def\@bals#1\eal{\begin{align*}#1\end{align*}}
\def\@bal#1\eal{\begin{align}#1\end{align}}
\newcommand{\ba}{\@ifstar{\@bas}{\@ba}}
\def\@bas#1\ea{\begin{equation*}\aligned#1\endaligned\end{equation*}}
\def\@ba#1\ea{\begin{equation}\aligned#1\endaligned\end{equation}}
\newcommand{\beq}{\@ifstar{\@beqs}{\@beq}}
\def\@beqs#1\eeq{\begin{equation*}#1\end{equation*}}
\def\@beq#1\eeq{\begin{equation}#1\end{equation}}
\newcommand{\beqa}{\@ifstar{\@beqas}{\@beqa}}
\def\@beqas#1\eeqa{\begin{eqnarray*}#1\end{eqnarray*}}
\def\@beqa#1\eeqa{\begin{eqnarray}#1\end{eqnarray}}
\makeatletter
\begin{document}
	
	\newcommand*{\ssp}{\scriptscriptstyle}
	\newcommand*{\gm}{\gamma}
	\newcommand*{\nm}{\nonumber}

\title{Impacts of (inverse) magnetic catalysis on screening masses of neutral pions and sigma mesons in hot and magnetized quark matter}

\author{Bing-kai Sheng}
\affiliation{Center for theoretical physics and College of Physics, Jilin University, Changchun 130012, P.R. China}
\affiliation{School of Fundamental Physics and Mathematical Sciences,
	Hangzhou Institute for Advanced Study, UCAS, Hangzhou, 310024, China}

\author{Xinyang Wang}
\email{wangxy@ujs.edu.cn}
\affiliation{ Department of Physics, Jiangsu University, Zhenjiang 212013, P.R. China}

\author{Lang Yu}
\email{yulang@jlu.edu.cn}
\affiliation{Center for theoretical physics and College of Physics, Jilin University, Changchun 130012, P.R. China}

\date{\today}	

\begin{abstract}
	\begin{spacing}{1.0}
We investigate the screening masses of neutral pions and sigma mesons in hot and magnetized quark matter in the framework of a two-flavor lattice-improved Nambu\textendash Jona-Lasinio (NJL) model with a magnetic field dependent coupling constant, which is determined by utilizing the results from lattice QCD simulations. Since such model can well reproduce inverse magnetic catalysis (IMC), by comparing with the standard NJL model, we systemically analyze the impacts of IMC on the temperature and magnetic field dependences of the longitudinal and transverse screening masses of the chiral partners, i.e. $\pi^0$ and $\sigma$ mesons, as well as the screening mass differences between them. 
Particularly, it is found that the $eB$
dependences of two alternative (pseudo)critical temperatures for the chiral transition defined by $\sigma-\pi^0$ meson screening mass differences are consistent with that defined by the quark condensate.  	 
	\end{spacing}
\end{abstract}	

\maketitle

\section{\label{sec:one}Introduction}
The properties of strong interactions, which is described by the theory of Quantum chromodynamics (QCD), in the presence of external magnetic fields has been extensively investigated in the past years (see, e.g., Refs.~\cite{Kharzeev:2015kna,Miransky:2015ava,Andersen:2014xxa} for recent reviews). This is because that strong magnetic fields are expected to exist in the early universe~\cite{Vachaspati:1991nm,Enqvist:1993np}, compact stars like magnetars~\cite{Duncan:1992hi}, and the non-central relativistic heavy ion collisions at RHIC and LHC~\cite{Skokov:2009qp,Voronyuk:2011jd, Bzdak:2011yy,Deng:2012pc}. In these physical situations, the magnitude of the
magnetic fields, ranging from $m^2_{\pi}$ to multiples of $m^2_{\pi}$, could be comparable with strong interactions ($eB\gtrsim \Lambda_{QCD}^2$),
so that the properties of strongly interacting matter will be significantly modified by such
strong magnetic fields. Particularly, there are a variety of new and intriguing phenomena induced by the interplay between magnetic fields and non-perturbative properties of QCD, for instance, chiral magnetic effect (CME)~\cite{Kharzeev:2007tn,Okorokov:2009bf,Fukushima:2008xe}, magnetic catalysis (MC)~\cite{Klevansky:1989vi,Klimenko:1990rh,Gusynin:1995nb,Shovkovy:2012zn}, inverse magnetic catalysis (IMC)~\cite{Bali:2011qj,Bali:2012zg} and vacuum superconductivity~\cite{Chernodub:2010qx,Chernodub:2011mc}, and so on. Thus, a deep and
comprehensive theoretical understanding of the dependence of QCD phase structure on the background magnetic field is desired.

One of important aspects regards the influence of the magnetic field on the chiral symmetry breaking of QCD. It is well known that the chiral condensate is the order parameter of the chiral phase transition in the chiral limit. And for physical quark masses, this transition
is changed from the second-order one to the crossover at zero baryon density, revealed by the Refs.~\cite{Pisarski:1983ms,Borsanyi:2010bp}, but the chiral condensate can still act as an approximate order parameter to exhibit the characteristic behavior of chiral symmetry breaking and restoration. When the external magnetic field is present, early lattice studies~\cite{Buividovich:2008wf,Braguta:2010ej,DElia:2010abb,DElia:2011koc,Ilgenfritz:2012fw}, as well as almost all low-energy effective models and theories of QCD~\cite{Klevansky:1989vi,Klimenko:1990rh,Gusynin:1995nb,Shovkovy:2012zn,Shushpanov:1997sf,
Agasian:1999sx,Alexandre:2000yf,Agasian:2001hv,Cohen:2007bt,Gatto:2010qs,Gatto:2010pt,Mizher:2010zb,
Kashiwa:2011js,Avancini:2012ee,Andersen:2012dz,Scherer:2012nn}, showed that the chiral condensate increases as the magnetic filed grows, a phenomenon called
magnetic catalysis~\cite{Klevansky:1989vi,Klimenko:1990rh,Gusynin:1995nb,Shovkovy:2012zn}, and correspondingly the pseudo-critical temperature $T_{pc}$ of the chiral phase transition increases with the magnetic field $B$. However, according to recent results of the lattice simulations by employing staggered quarks with physical quark masses, it is found that magnetic catalysis remains at low temperature, whereas at the temperature around $T_{pc}$ the chiral condensate would be surprisingly decreased by the magnetic field, which is called inverse magnetic catalysis~\cite{Bali:2011qj,Bali:2012zg}. As a result of such effect, the chiral pseudo-critical temperature is 
evidently reduced by the increasing of magnetic field~\cite{Bali:2011qj,Bali:2012zg}. since then, there have been a large number of studies~\cite{Fukushima:2012kc,Kojo:2012js,Bruckmann:2013oba,Chao:2013qpa,Fraga:2013ova,Ferreira:2014kpa,Farias:2014eca,Yu:2014sla,Andersen:2014oaa,Ferrer:2014qka,Providencia:2014txa,Farias:2016gmy,Mao:2016fha,Mamo:2015dea,Endrodi:2019whh,Endrodi:2019zrl,Tomiya:2019nym,He:2020fdi} trying to explore the underlying physics 
behind this puzzle problem. In particular, one possible idea is that by incorporating magnetic-dependent or thermo-magnetic-dependent coupling constants
in the NJL-type models~\cite{Ferreira:2014kpa,Farias:2014eca,Providencia:2014txa,Farias:2016gmy,Endrodi:2019whh}, IMC along with the decreasing behavior of $T_{pc}$ with $B$ can be reproduced to a great extent. And it might be considered as an indirect way of introducing the sea effect in the effective model descriptions, since the sea effect contribution, qualitatively speaking, is the main physical reason for the appearance of IMC, as argued by the lattice simulations in Ref.~\cite{Bruckmann:2013oba}. 

On the other hand, the response of the properties of mesons to $B$ will help to explore the phase structure of QCD in an
external magnetic field. For example, light mesons like pions are Nambu-Goldstone bosons corresponding to the chiral symmetry breaking, so the study on their properties is conducive to understanding the chiral phase transition under the magnetic field. Besides, for the chiral partners such as neutral pion and sigma mesons, their mass difference can be also considered as an order parameter to describe the behavior of the chiral crossover. For this reason, the pole masses of light mesons in the presence of magnetic fields have been widely evaluated in the framework of chiral perturbation theory~\cite{Andersen:2012zc}, the linear sigma model~\cite{Ayala:2018zat,Das:2019ehv,Ayala:2020dxs}, NJL model~\cite{Klevansky:1991ey,Fayazbakhsh:2012vr,Fayazbakhsh:2013cha,Avancini:2015ady,Avancini:2016fgq,Mao:2017wmq,
GomezDumm:2017jij,Wang:2017vtn,Liu:2018zag,Avancini:2018svs,Chaudhuri:2019lbw,Coppola:2019uyr,
Sheng:2020hge,Li:2020hlp,Xu:2020yag}, relativistic or non-relativistic constituent quark model~\cite{Orlovsky:2013gha,Simonov:2015xta,Kojo:2021gvm} and lattice QCD (LQCD) simulations~\cite{Hidaka:2012mz,Larina:2014sav,Luschevskaya:2014lga,Luschevskaya:2015bea,Luschevskaya:2015cko,Brandt:2015hnz,Bali:2017ian,Ding:2020jui,Ding:2020hxw}. In addition, another important effect of the magnetic field on mesons is that 
charged vector mesons are conjectured to condense for sufficiently strong magnetic fields~\cite{Chernodub:2010qx,Chernodub:2011mc}. The existence of the charged rho condensation
has been studied by a lot of work~\cite{Chernodub:2010qx,Chernodub:2011mc,Callebaut:2010mct,Ammon:2011je,Cai:2013pda,Frasca:2013kka,
Andreichikov:2013zba,Liu:2014uwa,Liu:2015pna,
Liu:2016vuw,Kawaguchi:2015gpt,Zhang:2016qrl,Ghosh:2016evc,Ghosh:2017rjo,Hidaka:2012mz,Larina:2014sav,Luschevskaya:2015bea,Bali:2017ian,Ding:2020jui}, and it is still an open question right now.

However, as we have mentioned in Ref.~\cite{Sheng:2020hge}, unlike the pole masses, the screening masses of light mesons at finite temperature under the magnetic field were studied in only a few trials~\cite{Fayazbakhsh:2012vr,Fayazbakhsh:2013cha,Wang:2017vtn,Sheng:2020hge}. In fact, the screening meson masses are also useful quantities for understanding
the properties of QCD, since light mesons play an important role in nuclear physics as mediators of nuclear or quark interactions and the range of force is determined by the inverse of screening mass. Especially, the screening mass difference between mesonic chiral partners is strongly related to chiral symmetry restoration of QCD. Namely, the screening masses of them become degenerate when the chiral symmetry gets restored. Hence, the temperature dependence of meson screening masses at vanishing magnetic field has been investigated in the NJL-type models~\cite{Kunihiro:1991hp,Florkowski:1993br,Ishii:2013kaa,Wang:2013wk,Ishii:2015ira}, in  holography QCD~\cite{Cao:2021tcr} and in LQCD~\cite{Cheng:2010fe,Maezawa:2013nxa,Kaczmarek:2013kva,Bazavov:2019www}. But when a magnetic field is applied, there were only several papers~\cite{Fayazbakhsh:2012vr,Fayazbakhsh:2013cha,Wang:2017vtn,Sheng:2020hge} that explored the influence of magnetic fields on the screening masses of mesons in the hot medium. Note that in our recent work~\cite{Sheng:2020hge}, we systematically presented the calculations of the screening masses for neutral pions under magnetic fields in the NJL model within full random phase approximation (RPA), and solved the limitations of the previous studies. Now we hope to extend our evaluations to the screening mass difference between neutral pions and sigma mesons
in the magnetic field, in order to analyze the effects of magnetic fields on the chiral phase transition in an alternative way. Furthermore, it is found in Refs.~\cite{Mao:2017wmq,Avancini:2018svs,Sheng:2020hge} that, the pole masses of neutral pions suffer a sudden jump at Mott transition temperature in the presence of an external magnetic field. This discontinuity may disturb the study for the $B$-dependent behavior of the chiral phase transition. Whereas, the temperature dependence of the screening masses for neutral pions has no jump behavior in the whole temperature range at finite magnetic field~\cite{Sheng:2020hge}. This implies that the mesonic screening masses may play a better role than pole masses in the investigations for the breaking or restoration of chiral symmetries under magnetic fields.     

In this work, we will concentrate on studying the impacts of IMC on the screening masses of neutral pion and sigma mesons, especially the mass difference between them, in hot and magnetized quark matter within the framework of a lattice-improved two-flavor NJL model, which can reproduce IMC by adopting a magnetic-dependent coupling constant~\cite{Bali:2011qj}. In this scenario, we phenomenologically determine the magnetic field dependence of the four-fermion interaction coupling constant utilizing the magnetic dependent constituent quark masses, which are inferred from the baryon mass spectrum as a function of the magnetic field by employing lattice simulations~\cite{Endrodi:2019whh}. Moreover, it should be emphasized that we choose the proper-time regularization scheme which can guarantee the sound velocities of mesons can not be larger than unity, according to the Ref~\cite{Sheng:2020hge}. And an infrared cut-off $\Lambda_{\ssp{IR}}$ is introduced by considering the color confinement so as to remove unphysical decay thresholds for sigma mesons into quarks and anti-quarks~\cite{Ebert:1996vx}. 

This paper is organized as follows. In Sec.~\ref{sec:two}, we will first introduce the two-flavor NJL model in the presence of an external magnetic field and show the gap equation in the mean field approximation. Next, in Sec.~\ref{sec:three}, the mesonic correlation function is derived by using
the RPA approach and the analytical expressions of the polarization functions for scalar-isoscalar 
and pseudoscalar-isovector channels are calculated in detail by using the propagators of quarks in 
the magnetic field. And then, in Sec.~\ref{sec:four}, after incorporating the lattice-improved NJL model with the magnetic-dependent coupling constants, we will show the corresponding numerical results and make some discussions. Finally, the summary and conclusions will be presented in Sec.~\ref{sec:five}.                    

\section{\label{sec:two}NJL model and the gap equation}

In the presence of an external electromagnetic field, the two-flavor NJL model~\cite{Nambu:1961tp,Nambu:1961fr} reads
\beq\label{Eq1}
\mathcal{L}_{NJL}=\bar{\psi}(i\gm^{\mu}D_{\mu}-\hat{m})\psi+G[(\bar{\psi}\psi)^2+(\bar{\psi}i\gm_{\ssp{5}}\bm{\tau}\psi)^2]~,
\eeq
where $D_{\mu}=\partial_{\mu}+i\hat{Q}eA^{ext}_{\mu}$ is the covariant derivative which couples quarks to an external $\mathrm{U(1)}$ gauge field $A^{ext}_{\mu}$, i.e., the electromagnetic field, and  $\hat{Q}=diag(Q_{u},Q_{d})=diag(2/3,-1/3)$ is the quark charge matrix in the two-flavor space. The current quark mass marix is $\hat{m}=diag(m_{u},m_{d})$ and explicitly brings chiral symmetry breaking. In this paper, the current masses of up and down quark are set to equal with each other, i.e., $m_{u}=m_{d}=m_0$ but the isospin symmetry is still broken by the external electromagnetic field. The capital $G$ is the coupling constant corresponding to the scalar and pseudo-scalar channels and $\bm{\tau}=(\tau_{1},\tau_{2},\tau_{3})$ are Pauli matrices in the two-flavor space. The fields of quarks are denoted by $\psi=(u,d)^{T}$. 

The constituent quark mass $m$ is determined by the gap equation
\beq\label{Eq2}
m=m_0-2G\left\langle\bar{\psi}\psi\right\rangle~,
\eeq     
which is derived by using the Hartree approximation~\cite{Klevansky:1992qe,Florkowski:1997pi}. Here, the quark (chiral) condensate $\left\langle\bar{\psi}\psi\right\rangle$ is defined by $\left\langle\bar{\psi}\psi\right\rangle\equiv-\mathrm{Tr}S(x,x)$ where $\mathrm{Tr}$ denotes the trace of the propagator of constituent quarks, i.e., $S(x,x')$ in flavor, color and spinor space. The propagator of constituent quarks is defined by
\beq\label{Eq3}
(i\gm^{\mu}D_{\mu}-m)S(x,x')=i\delta^{(4)}(x-x')~.
\eeq
In this paper, we investigate the effects of an external constant magnetic field on the screening masses of the mesons and thus choose the Landau gauge $A_{\mu}^{ext}=(0,0,-Bx,0)$ corresponding to the magnetic field $\bm{B}=(0,0,B)$ along the positive $z$ direction. The Lorentz symmetry is broken by the background magnetic field at zero temperature, i.e., $\mathrm{SO}(1,3) \to \mathrm{SO}(2)\otimes\mathrm{SO}(1,1)$. From Eq.~(\ref{Eq3}), we can obtain the analytical result of the propagator of constituent quarks in Minkowski space, i.e., $g_{\mu\nu}=diag(1,-1,-1,-1)$ and it reads
\beqa\label{Eq4}
S(x,x')=e^{i\Phi_{\ssp{f}}(\bm{r}_{\ssp{\perp}},\bm{r}_{\ssp{\perp}}')}\widetilde{S}(x-x')~,
\eeqa
where $\Phi_{\ssp{f}}(\bm{r}_{\ssp{\perp}},\bm{r}_{\ssp{\perp}}')=\dfrac{Q_{\ssp{f}}eB(x+x')(y-y')}{2}$ is the so-called Schwinger phase~\cite{PhysRev.82.664} and the index $f$ is the flavor index, i.e., $f=u,~d$ and $\bm{r}_{\ssp{\perp}}=(x,y)$. The translation invariant part of the propagator $\widetilde{S}(x-x')$ reads
\beq\label{Eq5}
\widetilde{S}(x-x')=\int\dfrac{d^4p}{(2\pi)^4}e^{-ip\cdot(x-x')}\widetilde{S}(p)~,
\eeq     
where
\bal\label{Eq6}
\widetilde{S}(p)&=\int_{0}^{\infty}ds\exp\bigg\{is\left[p_0^2-p_3^2-m^2+i\epsilon\right]-is\bm{p}_{\ssp{\perp}}^2  \nm
\\
&~~~\times\dfrac{tan(sQ_{\ssp{f}}eB)}{sQ_{\ssp{f}}eB}\bigg\}\bigg[\gm^{\mu}p_{\mu}+m+(p^1\gm^2-p^2\gm^1) \nm
\\
&~~~\times tan(sQ_{\ssp{f}}eB)\bigg]\bigg[1-\gm^1\gm^2tan(sQ_{\ssp{f}}eB)\bigg]~.
\eal
Here, $\bm{p}_{\ssp{\perp}}^2=p_{1}^2+p_{2}^2$.

The gap equation in vacuum can be expressed explicitly as follows by using the quark propagator of Eq.~(\ref{Eq4}),
\beqa\label{Eq7}
m=m_0+4Gm\mathrm{I}_{1,vac}(m^2)~,
\eeqa
where the integral $\mathrm{I}_{1,vac}(m^2)$ is defined by
\bal\label{Eq8}
\mathrm{I}_{1,vac}(m^2)&\equiv\dfrac{N_c}{8\pi^2}\sum_{f=u,d}\int_{0}^{\infty}ds\dfrac{e^{-m^2s}}{s}|Q_{\ssp{f}}eB| \nm
\\
&~~~\times coth(s|Q_{\ssp{f}}eB|)~.
\eal
At finite temperature, the rotation symmetry of space-time is further broken by the external heat reservoir, namely $\mathrm{SO}(2)\otimes\mathrm{SO}(1,1) \to \mathrm{SO}(2)$, which means that the system is merely invariant under $\mathrm{SO}(2)$ transformation in the plane which is vertical to the direction of the magnetic field. The sum over the Matsubara frequencies of fermions needs to be introduced to the integral of the quark momenta, i.e., $p^0 \to i\omega_{\ssp{l}}^{\ssp{F}}$ where $\omega_{\ssp{l}}^{\ssp{F}}=(2l+1)\pi T,~l=0,\pm 1,\pm 2 \cdots$, and the gap equation reads
\beq\label{Eq9}
m=m_0+4Gm\mathrm{I}_{1}(m^2)~,
\eeq
where the integral $\mathrm{I}_{1}(m^2)$ is defined as follows
\bal\label{Eq10}
\mathrm{I}_{1}(m^2)&\equiv\dfrac{N_c}{4\sqrt{\pi^3}}\sum_{f=u,d}\int_{0}^{\infty}ds\dfrac{e^{-m^2s}}{\sqrt{s}}\bigg[T\theta_{2}(0,e^{-4\pi^2T^2s})\bigg] \nm
\\
&~~~\times|Q_{\ssp{f}}eB|coth(s|Q_{\ssp{f}}eB|)~,
\eal
and $\theta_{2}(u,q)$ is the Jacobi theta function~\cite{book:63138}. 
Note that when the temperature $T$ goes to zero, $\mathrm{I}_{1}(m^2) \to \mathrm{I}_{1,vac}(m^2)$ is due to
\beq\label{Eq11}
\displaystyle\lim_{T \to 0}T\theta_{2}(0,e^{-4\pi^2T^2s})=\dfrac{1}{2\sqrt{\pi s}}~.
\eeq

As mentioned above, we employ the proper-time regularization scheme with the ultraviolet cut-off $\Lambda_{\ssp{UV}}$ and the infrared cut-off $\Lambda_{\ssp{IR}}$~\cite{Hellstern:1997nv,Ebert:1996vx,Bentz:2001vc} in the model. The regularized integrals in the gap equations at zero and finite temperature read
\bal\label{Eq12}
\mathrm{I}_{1,vac}^{\ssp{PT}}(m^2)&\equiv\dfrac{N_c}{8\pi^2}\sum_{f=u,d}
\int_{\frac{\ssp{1}}{\ssp{\Lambda}_{\ssp{UV}}^{\ssp{2}}}}^{\frac{\ssp{1}}{\ssp{\Lambda}_{\ssp{IR}}^{\ssp{2}}}}
ds\dfrac{e^{-m^2s}}{s}|Q_{\ssp{f}}eB| \nm
\\
&~~~\times coth(s|Q_{\ssp{f}}eB|)
\eal
and
\bal\label{Eq13}
\mathrm{I}_{1}^{\ssp{PT}}(m^2)&\equiv\dfrac{N_c}{4\sqrt{\pi^3}}\sum_{f=u,d}
\int_{\frac{\ssp{1}}{\ssp{\Lambda}_{\ssp{UV}}^{\ssp{2}}}}^{\frac{\ssp{1}}{\ssp{\Lambda}_{\ssp{IR}}^{\ssp{2}}}}
ds\dfrac{e^{-m^2s}}{\sqrt{s}}\bigg[T\theta_{2}(0,e^{-4\pi^2T^2s})\bigg] \nm
\\
&~~~\times|Q_{\ssp{f}}eB|coth(s|Q_{\ssp{f}}eB|),
\eal
respectively. Note that the gap equations at zero and finite temperature, i.e., Eq.~(\ref{Eq7}) and Eq.~(\ref{Eq9}), are in Euclid space, by performing the Wick rotation $s \to -is$.

\section{\label{sec:three}The screening masses of neutral pion and sigma meson}

We evaluate the screening masses of neutral pions and sigma mesons by following Ref.~\cite{Ishii:2013kaa}. Firstly, we consider the mesonic correlation function defined by
\beq\label{Eq14}
\eta_{\ssp{\xi\xi}}(x)\equiv\left\langle0\right|\mathrm{T}\left[J_{\ssp{\xi}}(x)J_{\ssp{\xi}}^{\ssp{\dagger}}(0)\right]\left|0\right\rangle~,
\eeq
where $\xi=\pi^0$ for neutral pion and $\xi=\sigma$ for sigma meson and capital T denotes the time-ordered product. Note that there is no mixing term considered here. The third component of the pseudoscalar isovector current is
\beq\label{Eq15}
J_{\ssp{\pi}^{\ssp{0}}}(x)=\bar{\psi}(x)i\gm_{\ssp{5}}\tau_{3}\psi(x)~,
\eeq 
and the scalar isoscalar current is
\beq\label{Eq16}
J_{\ssp{\sigma}}(x)=\bar{\psi}(x)\psi(x)-\left\langle\bar{\psi}(x)\psi(x)\right\rangle~.
\eeq
Then, the Fourier transformation of $\eta_{\ssp{\xi\xi}}(x)$ reads
\beqa\label{Eq17}
\chi_{\ssp{\xi\xi}}(k)=i\int d^4xe^{ik\cdot x}\left\langle0\right|\mathrm{T}\left[J_{\ssp{\xi}}(x)J_{\ssp{\xi}}^{\ssp{\dagger}}(0)\right]\left|0\right\rangle~.
\eeqa
The correlation function in momentum space, i.e., $\chi_{\ssp{\xi\xi}}(k)$, can be obtained by using the RPA method~\cite{Klevansky:1992qe,Florkowski:1997pi} and it is given by
\beq\label{Eq18}
\chi_{\ssp{\xi\xi}}(k)=\Pi_{\ssp{\xi}}(k)+2G\Pi_{\ssp{\xi}}(k)\chi_{\ssp{\xi\xi}}(k)~,
\eeq
where the one quark-loop polarization function $\Pi_{\ssp{\xi}}(k)$ is defined by
\beqa\label{Eq19}
\Pi_{\ssp{\xi}}(k)\equiv-i\int\dfrac{d^4p}{(2\pi)^4}\mathrm{Tr}\big[\Gamma_{\ssp{\xi}}\widetilde{S}(p)\Gamma_{\ssp{\xi}}\widetilde{S}(p-k)\big]~,
\eeqa
and $\Gamma_{\ssp{\pi}^{\ssp{0}}}=i\gm_{\ssp{5}}\tau_{3}$, $\Gamma_{\ssp{\sigma}}=\bm{1}$. Note that two Schwinger phases of the quark-antiquark pair cancel with each other in the neutral meson polarization functions.
Substituting Eq.~(\ref{Eq6}) into Eq.~(\ref{Eq19}) and completing the tedious calculations, we obtain the explicit expressions of the polarization functions at zero temperature as follows:
\begin{widetext} 
\bal\label{Eq20}
\Pi_{\ssp{\pi}^{\ssp{0}}}^{\ssp{vac}}(\bm{k}_{\ssp{\perp}}^2,\bm{k}_{\ssp{\parallel}}^2)&=\dfrac{N_c}{4\pi^2}\sum_{f=u,d}\int_{0}^{\infty}ds\int_{0}^{1}du\exp\Bigg[-m^2s -\dfrac{s(1-u^2)}{4}\bm{k}_{\ssp{\parallel}}^2-\dfrac{cos h(s|Q_{\ssp{f}}e B|)-cos h(s u|Q_{\ssp{f}}e B|)}{2|Q_{\ssp{f}}e B|sin h(s|Q_{\ssp{f}}e B|)}\bm{k}_{\ssp{\perp}}^2\Bigg] \nm
\\
&~~~\times\Bigg\{\bigg(m^2+\dfrac{1}{s}-\dfrac{1-u^2}{4}\bm{k}_{\ssp{\parallel}}^2\bigg)\dfrac{|Q_{\ssp{f}}e B|}{tanh(s|Q_{\ssp{f}}e B|)}-\dfrac{|Q_{\ssp{f}}e B|\left[cos h(s|Q_{\ssp{f}}e B|)-cos h(s u|Q_{\ssp{f}}e B|)\right]}{2sin h^3(s|Q_{\ssp{f}}e B|)}\bm{k}_{\ssp{\perp}}^2 \nm
\\
&~~~+\dfrac{|Q_{\ssp{f}}e B|^2}{sin h^2(s|Q_{\ssp{f}}e B|)} \Bigg\}~,
\eal
\bal\label{Eq21}
\Pi_{\ssp{\sigma}}^{\ssp{vac}}(\bm{k}_{\ssp{\perp}}^2,\bm{k}_{\ssp{\parallel}}^2)&=\dfrac{N_c}{4\pi^2}\sum_{f=u,d}\int_{0}^{\infty}ds\int_{0}^{1}du\exp\Bigg[-m^2s -\dfrac{s(1-u^2)}{4}\bm{k}_{\ssp{\parallel}}^2-\dfrac{cos h(s|Q_{\ssp{f}}e B|)-cos h(s u|Q_{\ssp{f}}e B|)}{2|Q_{\ssp{f}}e B|sin h(s|Q_{\ssp{f}}e B|)}\bm{k}_{\ssp{\perp}}^2\Bigg] \nm
\\
&~~~\times\Bigg\{\bigg(-m^2+\dfrac{1}{s}-\dfrac{1-u^2}{4}\bm{k}_{\ssp{\parallel}}^2\bigg)\dfrac{|Q_{\ssp{f}}e B|}{tanh(s|Q_{\ssp{f}}e B|)}-\dfrac{|Q_{\ssp{f}}e B|\left[cos h(s|Q_{\ssp{f}}e B|)-cos h(s u|Q_{\ssp{f}}e B|)\right]}{2sin h^3(s|Q_{\ssp{f}}e B|)}\bm{k}_{\ssp{\perp}}^2 \nm
\\
&~~~+\dfrac{|Q_{\ssp{f}}e B|^2}{sin h^2(s|Q_{\ssp{f}}e B|)} \Bigg\}~.
\eal
\end{widetext}
Here, $\bm{k}_{\ssp{\parallel}}^2=k_3^2+k_{4}^{2}$ and $\bm{k}_{\ssp{\perp}}^2=k_1^2+k_2^2$. Note that we have made the Wick rotation $s \to -is$ and $k^0 \to ik_4$.

At finite temperature, after making Matsubara frequency summation~\cite{Florkowski:1997pi,book:18639} we obtain the polarization functions as follows
\begin{widetext}
\bal\label{Eq22}
\Pi_{\ssp{\pi}^{\ssp{0}}}(\omega^{\ssp{B}}_{\ssp{m}},\bm{k}_{\ssp{\perp}}^2,k_3^2)&=\dfrac{N_cT}{4\sqrt{\pi^3}}\sum_{f=u,d}\sum_{l=-\infty}^{\infty}\int_{0}^{\infty}ds\int_{-1}^{1}du\sqrt{s}\exp\Bigg\{-s\bigg[(\omega_{\ssp{l}}^{\ssp{F}})^2+m^2\bigg]+s(1-u)\omega_{\ssp{l}}^{\ssp{F}}\omega^{\ssp{B}}_{\ssp{m}}-\dfrac{s}{2}(1-u)(\omega^{\ssp{B}}_{\ssp{m}})^2 \nm
\\
&~~~-\dfrac{s(1-u^2)}{4}k_3^2-\dfrac{cosh(s|Q_{\ssp{f}}eB|)-cosh(s u|Q_{\ssp{f}}eB|)}{2sinh(s|Q_{\ssp{f}}eB|)}\dfrac{\bm{k}_{\ssp{\perp}}^2}{|Q_{\ssp{f}}eB|}\Bigg\}\Bigg\{
\bigg[m^2+\dfrac{1}{2s}-i\omega_{\ssp{l}}^{\ssp{F}}(i\omega_{\ssp{l}}^{\ssp{F}}-i\omega^{\ssp{B}}_{\ssp{m}})\nm
\\
&~~~-\dfrac{1-u^2}{4}k_3^2\bigg]\dfrac{|Q_{\ssp{f}}eB|}{tanh(s|Q_{\ssp{f}}eB|)}+\dfrac{|Q_{\ssp{f}}eB|^2}{sinh^2(s|Q_{\ssp{f}}eB|)}-\dfrac{|Q_{\ssp{f}}eB|\left[cosh(s|Q_{\ssp{f}}eB|)-cosh(su|Q_{\ssp{f}}eB|)\right]}{2sinh^3(s|Q_{\ssp{f}}eB|)}\bm{k}_{\ssp{\perp}}^2\Bigg\}~,
\eal
\bal\label{Eq23}
\Pi_{\ssp{\sigma}}(\omega^{\ssp{B}}_{\ssp{m}},\bm{k}_{\ssp{\perp}}^2,k_3^2)&=\dfrac{N_cT}{4\sqrt{\pi^3}}\sum_{f=u,d}\sum_{l=-\infty}^{\infty}\int_{0}^{\infty}ds\int_{-1}^{1}du\sqrt{s}\exp\Bigg\{-s\bigg[(\omega_{\ssp{l}}^{\ssp{F}})^2+m^2\bigg]+s(1-u)\omega_{\ssp{l}}^{\ssp{F}}\omega^{\ssp{B}}_{\ssp{m}}-\dfrac{s}{2}(1-u)(\omega^{\ssp{B}}_{\ssp{m}})^2 \nm
\\
&~~~-\dfrac{s(1-u^2)}{4}k_3^2-\dfrac{cosh(s|Q_{\ssp{f}}eB|)-cosh(s u|Q_{\ssp{f}}eB|)}{2sinh(s|Q_{\ssp{f}}eB|)}\dfrac{\bm{k}_{\ssp{\perp}}^2}{|Q_{\ssp{f}}eB|}\Bigg\}\Bigg\{
\bigg[-m^2+\dfrac{1}{2s}-i\omega_{\ssp{l}}^{\ssp{F}}(i\omega_{\ssp{l}}^{\ssp{F}}-i\omega^{\ssp{B}}_{\ssp{m}})\nm
\\
&~~~-\dfrac{1-u^2}{4}k_3^2\bigg]\dfrac{|Q_{\ssp{f}}eB|}{tanh(s|Q_{\ssp{f}}eB|)}+\dfrac{|Q_{\ssp{f}}eB|^2}{sinh^2(s|Q_{\ssp{f}}eB|)}-\dfrac{|Q_{\ssp{f}}eB|\left[cosh(s|Q_{\ssp{f}}eB|)-cosh(su|Q_{\ssp{f}}eB|)\right]}{2sinh^3(s|Q_{\ssp{f}}eB|)}\bm{k}_{\ssp{\perp}}^2\Bigg\}~.
\eal
\end{widetext}
Since we evaluate the screening masses, the Matsubara frequences of bosons (mesons) $\omega_{\ssp{m}}^{\ssp{B}}=2m\pi T$ with $m=0,\pm 1, \pm 2 \cdots$, which emerge in Eq.~(\ref{Eq22}) and Eq.~(\ref{Eq23}) should be vanished, namely $k_4=0$. As shown in the Ref.~\cite{Ishii:2013kaa}, the screening masses of mesons in a heat reservoir but with vanishing magnetic field can be calculated by making use of the 
dispersion relation obtained from Eq.~(\ref{Eq18}), i.e.,
\beq\label{Eq24}
\left[1-2G\Pi_{\ssp{\xi}}(k_4=0,\bm{k}^2)\right]|_{\bm{k}^2=-m^2_{\ssp{\xi},scr}}=0,~\rm{for}~B\rightarrow 0
\eeq
where $\bm{k}=(k^1,k^2,k^3)$.
In the presence of an external magnetic field, analogously, the longitudinal screening masses $m_{\ssp{\xi},scr,\ssp{\parallel}}$ and the transverse screening masses $m_{\ssp{\xi},scr,\ssp{\perp}}$ of the mesons are determined by 
\beq\label{Eq25}
\left[1-2G\Pi_{\ssp{\xi}}^{\ssp{vac}}(0,k_3^2)\right]|_{k_3^2=-m^2_{\ssp{\xi},scr,\ssp{\parallel}}}=0~
\eeq
and
\beq\label{Eq26}
\left[1-2G\Pi_{\ssp{\xi}}^{\ssp{vac}}(\bm{k}_{\ssp{\perp}}^2,0)\right]|_{\bm{k}_{\ssp{\perp}}^2=-m^2_{\ssp{\xi},scr,\ssp{\perp}}}=0
\eeq
for $T=0$, and by
\beq\label{Eq27}
\left[1-2G\Pi_{\ssp{\xi}}(0,0,k_3^2)\right]|_{k_3^2=-m^2_{\ssp{\xi},scr,\ssp{\parallel}}}=0~
\eeq
and
\beq\label{Eq28}
\left[1-2G\Pi_{\ssp{\xi}}(0,\bm{k}_{\ssp{\perp}}^2,0)\right]|_{\bm{k}_{\ssp{\perp}}^2=-m^2_{\ssp{\xi},scr,\ssp{\perp}}}=0
\eeq
for $T\neq 0$.

And then, with the help of the method for regularizing the polarization functions in Ref.~\cite{Klevansky:1991ey}, we make an extension  from the Pauli-Villars regularization scheme to the proper-time regularization scheme with the infrared cut-off $\Lambda_{\ssp{IR}}$. The explicit regularized expressions of the polarization functions at zero and finite temperature are changed to the forms as follows, respectively,
\bal\label{Eq29}
\Pi_{\ssp{\xi},vac}^{\ssp{PT}}(\bm{k}_{\ssp{\perp}}^2,\bm{k}_{3}^2)&=2\mathrm{I}_{1,vac}^{\ssp{PT}}(m^2)
+(\bm{k}_{\ssp{\parallel}}^2+4m^2\epsilon_{\ssp{\xi}})\mathrm{I}_{2,vac,\ssp{\parallel}}^{\ssp{PT}}
(\bm{k}_{\ssp{\perp}}^2,\bm{k}_{3}^2) \nm
\\
&~~~+\bm{k}_{\ssp{\perp}}^2\mathrm{I}_{2,vac,\ssp{\perp}}^{\ssp{PT}}(\bm{k}_{\ssp{\perp}}^2,\bm{k}_{3}^2)
\eal
and
\bal\label{Eq30}
\Pi_{\ssp{\xi}}^{\ssp{PT}}(0,\bm{k}_{\ssp{\perp}}^2,k_3^2)&=2\mathrm{I}_{1}^{\ssp{PT}}(m^2)
+(k_3^2+4m^2\epsilon_{\ssp{\xi}})\mathrm{I}_{2,\ssp{\parallel}}^{\ssp{PT}}(0,\bm{k}_{\ssp{\perp}}^2,k_3^2) \nm
\\
&~~~+\bm{k}_{\ssp{\perp}}^2\mathrm{I}_{2,\ssp{\perp}}^{\ssp{PT}}(0,\bm{k}_{\ssp{\perp}}^2,k_3^2)~,
\eal
where $\epsilon_{\ssp{\xi}}$ is defined by
\beq\label{Eq31}
\epsilon_{\ssp{\xi}}\equiv
\begin{cases}
	1,~&\xi=\sigma \\
	0,~&\xi=\pi^0
\end{cases}
\eeq
and
\begin{widetext}
\bal\label{Eq32}
\mathrm{I}_{2,vac,\ssp{\parallel}}^{\ssp{PT}}(\bm{k}_{\ssp{\perp}}^2,\bm{k}_{3}^2)&=-\dfrac{N_c}{8\pi^2}\sum_{f=u,d}\int_{\frac{\ssp{1}}{\ssp{\Lambda}_{\ssp{UV}}^{\ssp{2}}}}^{\frac{\ssp{1}}{\ssp{\Lambda}_{\ssp{IR}}^{\ssp{2}}}}ds\int_{0}^{1}du\exp\Bigg[-m^2s -\dfrac{s(1-u^2)}{4}\bm{k}_{3}^2-\dfrac{cos h(s|Q_{\ssp{f}}e B|)-cos h(s u|Q_{\ssp{f}}e B|)}{2|Q_{\ssp{f}}e B|sin h(s|Q_{\ssp{f}}e B|)}\bm{k}_{\ssp{\perp}}^2\Bigg] \nm
\\
&~~~\times\dfrac{|Q_{\ssp{f}}eB|}{tanh(s|Q_{\ssp{f}}eB|)}~,
\eal
\bal\label{Eq33}
\mathrm{I}_{2,vac,\ssp{\perp}}^{\ssp{PT}}(\bm{k}_{\ssp{\perp}}^2,\bm{k}_{3}^2)&=-\dfrac{N_c}{8\pi^2}\sum_{f=u,d}
\int_{\frac{\ssp{1}}{\ssp{\Lambda}_{\ssp{UV}}^{\ssp{2}}}}^{\frac{\ssp{1}}{\ssp{\Lambda}_{\ssp{IR}}^{\ssp{2}}}}du\exp\Bigg[-m^2s -\dfrac{s(1-u^2)}{4}\bm{k}_{3}^2
-\dfrac{cos h(s|Q_{\ssp{f}}e B|)-cos h(s u|Q_{\ssp{f}}e B|)}{2|Q_{\ssp{f}}e B|sin h(s|Q_{\ssp{f}}e B|)}\bm{k}_{\ssp{\perp}}^2\Bigg] \nm
\\
&~~~\times\dfrac{|Q_{\ssp{f}}eB|cosh(su|Q_{\ssp{f}}eB|)}{sinh(s|Q_{\ssp{f}}eB|)}~,
\eal
\bal\label{Eq34}
\mathrm{I}_{2,\ssp{\parallel}}^{\ssp{PT}}(0,\bm{k}_{\ssp{\perp}}^2,k_3^2)&=-\dfrac{N_c}{4\sqrt{\pi^3}}\sum_{f=u,d}\int_{\frac{\ssp{1}}{\ssp{\Lambda}_{\ssp{UV}}^{\ssp{2}}}}^{\frac{\ssp{1}}{\ssp{\Lambda}_{\ssp{IR}}^{\ssp{2}}}}\int_{0}^{1}du\sqrt{s}\exp\Bigg[-m^2s -\dfrac{s(1-u^2)}{4}k_3^2-\dfrac{cos h(s|Q_{\ssp{f}}e B|)-cos h(s u|Q_{\ssp{f}}e B|)}{2|Q_{\ssp{f}}e B|sin h(s|Q_{\ssp{f}}e B|)}\bm{k}_{\ssp{\perp}}^2\Bigg] \nm
\\
&~~~\times\bigg[T\theta_{2}(0,e^{-4\pi^2T^2s})\bigg]\dfrac{|Q_{\ssp{f}}eB|}{tanh(s|Q_{\ssp{f}}eB|)}~,
\eal
and
\bal\label{Eq35}
\mathrm{I}_{2,\ssp{\perp}}^{\ssp{PT}}(0,\bm{k}_{\ssp{\perp}}^2,k_3^2)&=-\dfrac{N_c}{4\sqrt{\pi^3}}\sum_{f=u,d}
\int_{\frac{\ssp{1}}{\ssp{\Lambda}_{\ssp{UV}}^{\ssp{2}}}}^{\frac{\ssp{1}}{\ssp{\Lambda}_{\ssp{IR}}^{\ssp{2}}}}\int_{0}^{1}du\sqrt{s}\exp\Bigg[-m^2s -\dfrac{s(1-u^2)}{4}k_3^2-\dfrac{cos h(s|Q_{\ssp{f}}e B|)-cos h(s u|Q_{\ssp{f}}e B|)}{2|Q_{\ssp{f}}e B|sin h(s|Q_{\ssp{f}}e B|)}\bm{k}_{\ssp{\perp}}^2\Bigg] \nm
\\
&~~~\times\bigg[T\theta_{2}(0,e^{-4\pi^2T^2s})\bigg]\dfrac{|Q_{\ssp{f}}eB|cosh(su|Q_{\ssp{f}}eB|)}{sinh(s|Q_{\ssp{f}}eB|)}~.
\eal	
\end{widetext}

It is easy to find out the anisotropy between the longitudinal and transverse directions in terms of Eq.~(\ref{Eq29}) and Eq.~(\ref{Eq30}), since $\mathrm{I}_{2,vac,\ssp{\parallel}}\neq\mathrm{I}_{2,vac,\ssp{\perp}}$ and $\mathrm{I}_{2,\ssp{\parallel}}\neq\mathrm{I}_{2,\ssp{\perp}}$ at nonzero magnetic fields. And thus the longitudinal screening masses are different from the transverse screening ones, i.e., $m_{\ssp{\xi},scr,\ssp{\parallel}}\neq m_{\ssp{\xi},scr,\ssp{\perp}}$, when the magnetic field is present. But when $eB\to 0$, we have $\displaystyle\lim_{eB\to 0}\mathrm{I}_{2,vac,\ssp{\parallel}}=\lim_{eB\to 0}\mathrm{I}_{2,vac,\ssp{\perp}}$ at $T=0$ and $\displaystyle\lim_{eB\to 0}\mathrm{I}_{2,\ssp{\parallel}}=\lim_{eB\to 0}\mathrm{I}_{2,\ssp{\perp}}$ at $T\neq 0$. It means that the Lorentz symmetry broken by the magnetic field is going to be restored, i.e., $\mathrm{SO}(2)\otimes\mathrm{SO}(1,1)\to\mathrm{SO}(1,3)$ at zero temperature and $\mathrm{SO}(2)\to\mathrm{SO}(3)$ at finite temperature. Correspondingly, the relative index of refraction of the medium~\cite{Sheng:2020hge} becomes unity (i.e. $\frac{n_{\ssp{\xi},\ssp{\parallel}}}{n_{\ssp{\xi},\ssp{\perp}}}=\frac{m_{\ssp{\xi},scr,\ssp{\parallel}}}{m_{\ssp{\xi},scr,\ssp{\perp}}}=1$) in the absence of the magnetic field, whether the system is in heat reservoir or not.

\section{\label{sec:four}Numerical results}
\subsection{\label{sec:four,one}The $B$ dependent coupling constant $G$}

As mentioned above, it is an effective and convenient way to employ a magnetic field dependent four-fermion coupling constant in the NJL-type models, in order to incorporate inverse magnetic catalysis at high temperatures. Hence, by following Ref.~\cite{Endrodi:2019whh}, we determine the magnetic field dependence of the coupling constant $G$ by using the magnetic field dependent constituent quark masses inferred from the baryon masses in the magnetic fields obtained by lattice simulations. 

\begin{figure}[htbp]
\includegraphics[scale=0.27]{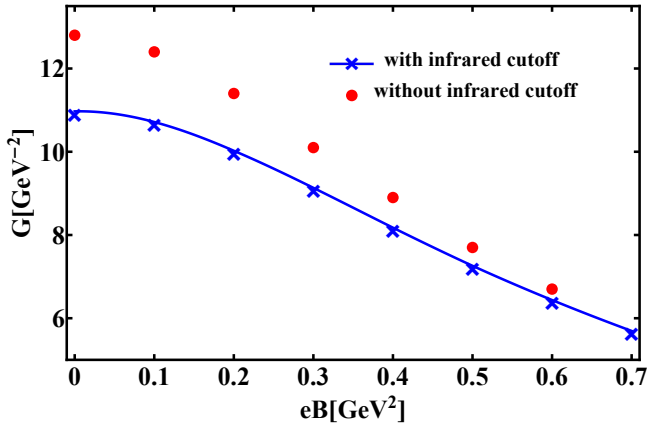}
\caption{\label{fig1}The coupling constant $G$ as a function of $eB$ compared with the results in Ref.~\cite{Endrodi:2019whh}. The blue line denotes the results in this paper and the red dots denote the results in Ref.~\cite{Endrodi:2019whh}.}	
\end{figure}

Specifically, we first fix the model parameters including the current quark mass $m_0$ and the ultraviolet cutoff $\Lambda_{\ssp{UV}}$ by setting the predictions of the model for the pion pole mass $m_{\ssp{\pi}}$ and for its decay constant $f_{\ssp{\pi}}$ to their physical values, i.e., 138MeV and 93MeV, respectively. And then, to fix the magnetic-dependent coupling constant $G(eB)$, we set the constituent quark mass at each $B$ to take the value inferred from the first-principles input of the baryon masses. Note that in this paper, we introduce an infrared cutoff that provides the quark confinement to eliminate unphysical quark-antiquark thresholds for mesons. And the infrared cutoff $\Lambda_{\ssp{IR}}=240\text{MeV}$ which is approximately equal to $\Lambda_{\ssp{QCD}}$. The results of the fixed model parameters are $m_0=11.8\text{MeV}$, $\Lambda_{\ssp{UV}}=708\text{MeV}$ and $G(T=B=0)\Lambda_{\ssp{UV}}^2=5.5$, and the values of the magnetic field dependent four-fermion coupling constant $G(eB)$, as well as the constituent quark masses used to fix them, are given in Table~\ref{tb1}. Moreover, the curve of the coupling constant $G$ as a function of $eB$ is also shown in Fig.~\ref{fig1}, in comparison with that in Ref.~\cite{Endrodi:2019whh}. Clearly, the coupling constant in our model decreases with
increasing magnetic field, which is qualitatively in agreement with the results in Ref.~\cite{Endrodi:2019whh}. And we need to emphasize that although the authors in Ref.~\cite{Endrodi:2019whh} worked in the Polyakov loop-extended NJL model (PNJL), they solely employed the input at zero temperature when fixing the magnetic field dependence of $G$, and thus the Polyakov loop contributed nothing.

\begin{table*}
\begin{ruledtabular}
\begin{tabular}{lcccccccc}
$\text{eB}~[\text{GeV}^2]$ & 0.0 & 0.1 & 0.2 & 0.3 & 0.4 & 0.5 & 0.6 & 0.7\\
\colrule
$\text{m}^2~[\text{GeV}^2]$ & 0.097 & 0.096 & 0.094 & 0.091 & 0.087 & 0.083 & 0.079 & 0.074\\
\colrule
$\text{G}~[\text{GeV}^{-2}]$ & 10.97 & 10.72 & 10.02 & 9.14 & 8.18 & 7.26 & 6.44 & 5.69\\
\end{tabular}
\end{ruledtabular}
\caption{\label{tb1}The values of the magnetic field dependent coupling constant and the constituent quark masses used to fix them.}
\end{table*} 
     
\subsection{\label{sec:four,two}The solution of the gap equation and the pseudo-critical temperature $T_{pc}$}

\begin{figure}[htbp]
	\includegraphics[scale=0.256]{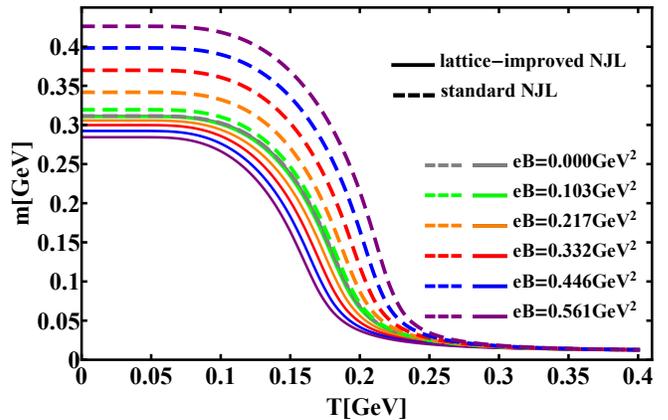}
	\caption{\label{fig2}The constituent quark mass as a function of temperature at different $eB$ compared with the results of standard NJL model.}	
\end{figure}

\begin{figure}[htbp]
	\includegraphics[scale=0.250]{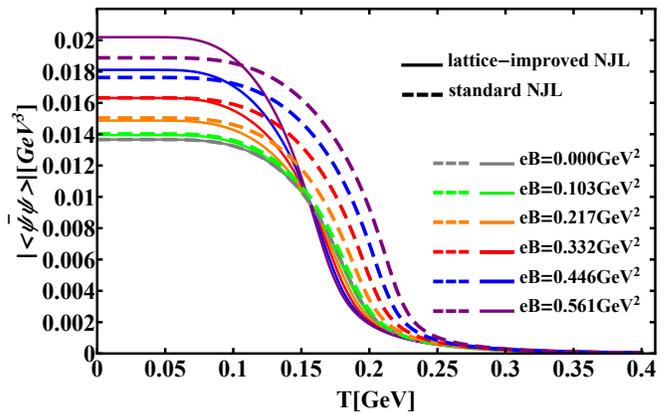}
	\caption{\label{fig3}The absolute value of the quark condensate as a function of temperature at different $eB$ compared with the standard NJL model.}
\end{figure}

In this subsection, we show the results of the constituent quark mass and the quark condensate as functions of temperature at different values of $eB$ in the framework of both the standard NJL model and the lattice-improved NJL model, as shown in Fig.~\ref{fig2} and Fig.~\ref{fig3}, respectively. It is found in Fig.~\ref{fig2} that, at a certain fixed temperature value, the constituent quark mass $m$ decreases with $eB$ in the lattice-improved NJL model with the magnetic-dependent coupling constant, while the situation is on the contrary in the standard NJL model with a constant $ G(B=0)$.

On the other hand, as depicted by Fig.~\ref{fig3}, in the standard NJL model, the quark condensate increases with the magnetic field at any temperatures, which is just referred to as magnetic catalysis. However, in the case of the lattice-improved NJL model, the quark condensate curves display magnetic catalysis only at low temperatures but show inverse magnetic catalysis at sufficiently high temperatures around the transition, which is consistent with the lattice results in Ref.~\cite{Bali:2012zg}. Note that the same values of $eB$ as Ref.~\cite{Endrodi:2019whh} are chosen for convenience in our calculations when we evaluate the constituent quark mass and the quark condensate.    

\begin{figure}[htbp]
	\includegraphics[scale=0.258]{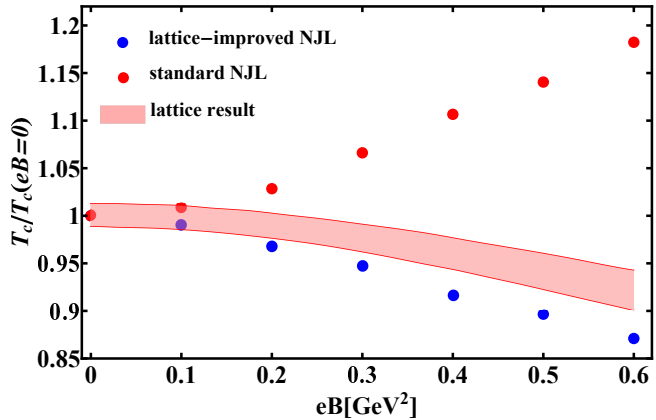}
	\caption{\label{fig4}The pseudo-critical temperature scaled by its $eB=0$ value as a function of the magnetic field compared with the results from lattice simulations~\cite{Bali:2011qj}.}	
\end{figure}

Using the results of the quark condensate in Fig.~\ref{fig3}, we plot the pseudo-critical temperature $T_{pc}$, defined by the location of the inflection points of the quark condensate curves, as a function of the magnetic field strength in Fig.~\ref{fig4}. It is obvious that our numerical results of $T_{pc}(B)$ in the lattice-improved NJL model is qualitatively in good agreement with the lattice results in Ref.~\cite{Bali:2011qj} as opposed to the standard NJL model. As shown in Ref.~\cite{Endrodi:2019whh},
\begin{figure*}[htbp]
	\includegraphics[scale=0.25]{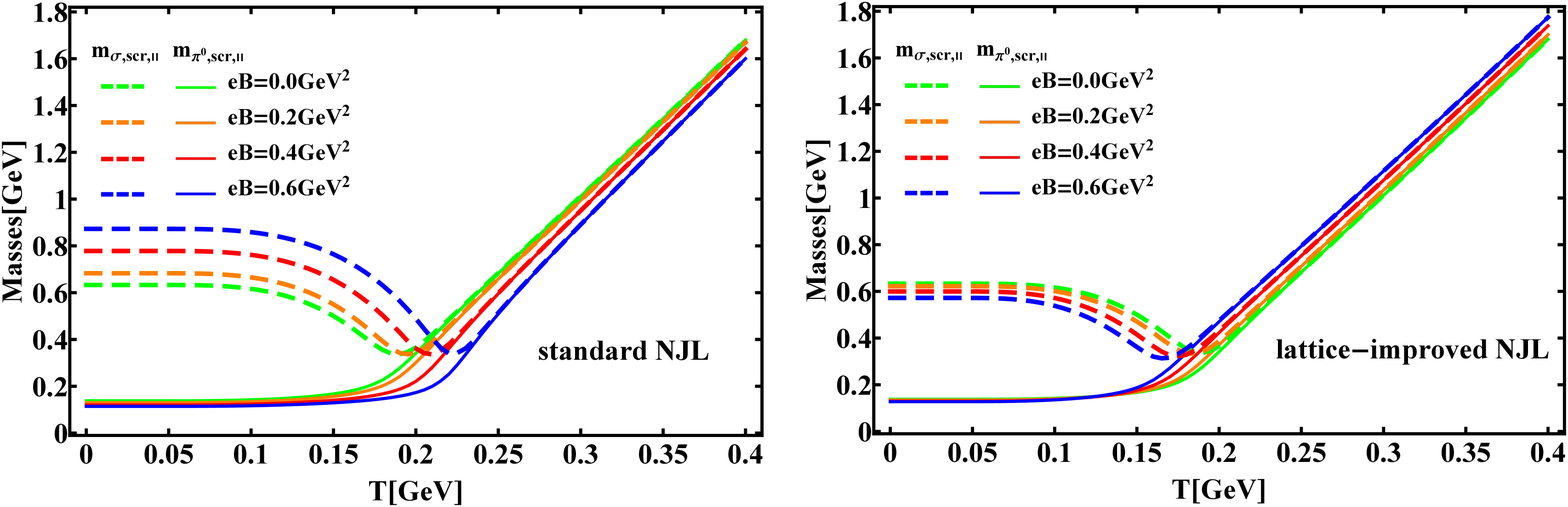}
	\caption{\label{fig5}Left panel: The longitudinal meson screening masses as functions of temperature at fixed $eB=0.0$, $0.2$, $0.4$ and $0.6$ $\rm{GeV^2}$ in the standard NJL model. Right panel: The same quantities in the lattice-improved NJL model.}
\end{figure*}
\begin{figure*}[htbp]
	\includegraphics[scale=0.25]{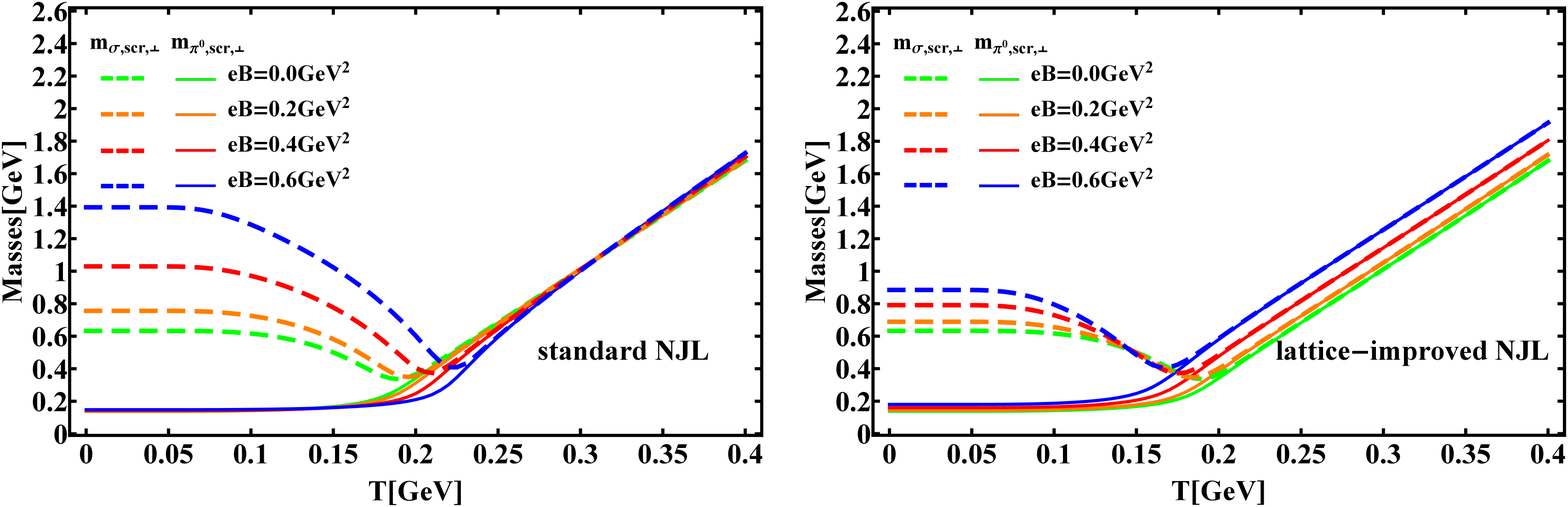}
	\caption{\label{fig6}Left panel: The transverse meson screening masses as functions of temperature at fixed $eB=0.0$, $0.2$, $0.4$ and $0.6$ $\rm{GeV^2}$ in the standard NJL model. Right panel: The same quantities in the lattice-improved NJL model.}
\end{figure*}
the authors actually used the PNJL model with magnetic field dependent $G$ to reproduce the IMC successfully. Our numerical results in the lattice-improved NJL model without the Polyakov loop are still consistent with those in Ref.~\cite{Endrodi:2019whh}. It implies that the key point for introducing IMC in the NJL-type models is the magnetic field dependent four-fermion coupling constant and the Polyakov loop is not essential. For this reason, our robust lattice-improved NJL model is qualified to investigate the effects of the IMC on the screening masses of mesons.

\subsection{\label{sec:four,three}The screening masses of neutral pion and sigma meson}
\subsubsection{\label{sec:four,three,one}Results at fixed $eB$}

The screening masses of neutral pions and sigma mesons can be evaluated in terms of Eqs.~(\ref{Eq25}), (\ref{Eq26}), (\ref{Eq27}) and (\ref{Eq28}). Now we begin the discussion with the results of their screening masses as functions of temperature at different values of $eB$ in the cases of the standard NJL model and the lattice-improved NJL model.

For the standard NJL model, the temperature dependences of the longitudinal and transverse screening masses for $\pi^0$ and $\sigma$ mesons at fixed $eB=0.0$, $0.2$, $0.4$ and $0.6$ $\rm{GeV^2}$ are shown in the left panels of Fig.~\ref{fig5} and Fig.~\ref{fig6}, respectively. In general, the curves of longitudinal screening masses show similar behavior with those of transverse ones. At low temperatures ($T\lesssim150\text{MeV}$), the longitudinal and transverse screening masses of both $\pi^0$ and $\sigma$ mesons hardly change with the increase of temperature. As the temperature becomes higher and higher, the longitudinal and transverse screening masses of neutral pions begin to increase with the increasing temperature $T$, while both screening masses of sigma mesons first decrease and then increase as $T$ is growing. Especially, at a certain temperature, defined by $T_{\chi}^{\parallel}$ (or $T_{\chi}^{\perp}$), the longitudinal (or transverse) screening masses of neutral pions and sigma mesons merge with each other. This phenomenon implies that the chiral symmetry is restored at $T_{\chi}^{\parallel}$ (or $T_{\chi}^{\perp}$) because neutral pion and sigma mesons are chiral partners. And it is shown that the merging temperature, either $T_{\chi}^{\parallel}$ or $T_{\chi}^{\perp}$, is enhanced by the magnetic field.       

As for our lattice-improved NJL model, the corresponding numerical results of the meson screening masses are shown in the right panels of Fig.~\ref{fig5} and Fig.~\ref{fig6}. By comparing with the results in the standard NJL model, we could examine the effects of the IMC on the meson screening masses. First, as shown by Fig.~\ref{fig5}, it is evident that  $m_{\ssp{\sigma},scr,\ssp{\parallel}}$ obtained in the lattice-improved NJL model is reduced by the magnetic field at low temperatures but is changed to increase with $B$ at high temperatures, in contrast to the standard NJL. Second, as regard to $m_{\ssp{\sigma},scr,\ssp{\perp}}$, as shown in Fig.~\ref{fig6}, it remains enhanced by the magnetic field at low temperatures in the lattice-improved NJL model, although the extent of the increment is less than that in the standard NJL model. However, while $T$ is around $T_{\chi}^{\perp}$, $m_{\ssp{\sigma},scr,\ssp{\perp}}$ acquired by the lattice-improved NJL model becomes increase with $B$, similar\ to $m_{\ssp{\sigma},scr,\ssp{\parallel}}$. Third, what's more important, when approaching $T_{\chi}^{\parallel}$ ($T_{\chi}^{\perp}$), the results of the lattice-improved NJL model show that, the larger the strength of the magnetic field, the lower the temperature where either $m_{\ssp{\sigma},scr,\ssp{\parallel}}$ ($m_{\ssp{\sigma},scr,\ssp{\perp}}$) or $m_{\ssp{\pi^0},scr,\ssp{\parallel}}$ ($m_{\ssp{\pi^0},scr,\ssp{\perp}}$) starts to increase with $T$ remarkably, as opposed to the results in the standard NJL. Consequently, the location where the two curves of the screening masses of neutral pions and sigma mesons merge together moves towards the vertical axis with the enhancement of the magnetic field. That is to say, $T_{\chi}^{\parallel}$ ($T_{\chi}^{\perp}$) is reduced by the increase of the magnetic field strength in the case of our lattice-improved NJL model. In addition, it is obvious that in the high temperature limit, the curves of $m_{\ssp{\sigma},scr,\ssp{\perp}}$ (or $m_{\ssp{\pi^0},scr,\ssp{\perp}}$) at different values of $eB$ in the standard NJL tends to approach each other as $T$ grows, while those curves in the lattice-improved NJL model keep increasing with $T$ separately and parallelly .

\begin{figure}[htbp]
	\includegraphics[scale=0.2465]{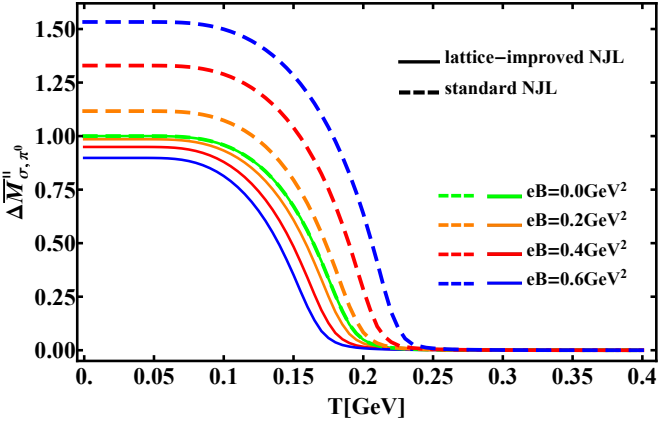}
	\caption{\label{fig7}The normalized longitudinal screening mass difference  $\Delta\overline{M}_{\sigma,\pi^0}^{\ssp{\parallel}}$ as a function of temperature at fixed $eB=0.0$, $0.2$, $0.4$ and $0.6$ $\rm{GeV^2}$.}
\end{figure}

\begin{figure}[htbp]
	\includegraphics[scale=0.2495]{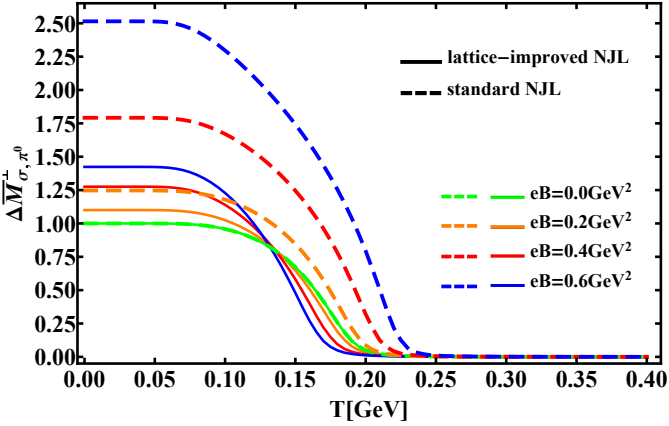}
	\caption{\label{fig8}The normalized transverse screening mass difference  $\Delta\overline{M}_{\sigma,\pi^0}^{\ssp{\perp}}$ as a function of temperature at fixed 
		$eB=0.0$, $0.2$, $0.4$ and $0.6$ $\rm{GeV^2}$.}
\end{figure}

Since the neutral pions and the sigma mesons are chiral partners with each other, we could introduce their screening mass difference to investigate the chiral phase transition as an alternative of the chiral condensate, as we have discussed above. Explicitly, the normalized screening mass difference between $\pi^0$ and $\sigma$ mesons in the longitudinal and the transverse directions are defined by  
\beq\label{Eq39}
\Delta\overline{M}_{\sigma,\pi^0}^{\ssp{\parallel}}(B,T)=\dfrac{\Delta{m}_{\sigma,\pi^0}^{\ssp{\parallel}}(B,T)}
{\Delta{m}_{\sigma,\pi^0}^{\ssp{\parallel}}(0,0)}
\eeq
and
\beq\label{Eq40}
\Delta\overline{M}_{\sigma,\pi^0}^{\ssp{\perp}}(B,T)=\dfrac{\Delta{m}_{\sigma,\pi^0}^{\ssp{\perp}}(B,T)}
{\Delta{m}_{\sigma,\pi^0}^{\ssp{\perp}}(0,0)},
\eeq
respectively, where $\Delta{m}_{\sigma,\pi^0}^{\ssp{\parallel}}=m_{\ssp{\sigma},scr,\ssp{\parallel}}-m_{\ssp{\pi^0},scr,\ssp{\parallel}}$
and $\Delta{m}_{\sigma,\pi^0}^{\ssp{\perp}}=m_{\ssp{\sigma},scr,\ssp{\perp}}-m_{\ssp{\pi^0},scr,\ssp{\perp}}$. And in Fig.~\ref{fig7} and Fig.~\ref{fig8}, we display the normalized difference $\Delta\overline{M}_{\sigma,\pi^0}^{\ssp{\parallel}}$ and $\Delta\overline{M}_{\sigma,\pi^0}^{\ssp{\perp}}$ as functions of temperature at $eB=0.0$, $0.2$, $0.4$ and $0.6$ $\rm{GeV}^2$.

It is shown that, compared with Fig.~\ref{fig2} and Fig.~\ref{fig3}, the temperature dependences of $\Delta\overline{M}_{\sigma,\pi^0}^{\ssp{\parallel}}$ and $\Delta\overline{M}_{\sigma,\pi^0}^{\ssp{\perp}}$ at different fixed values of $eB$ are analogous to those of the consituent quark mass and the chiral condensate respectively, no matter in the standard NJL model or the lattice-improved NJL model. Concretely, at low temperatures ($T\lesssim100\text{MeV}$), $\Delta\overline{M}_{\sigma,\pi^0}^{\ssp{\parallel}}$ or $\Delta\overline{M}_{\sigma,\pi^0}^{\ssp{\perp}}$ is nearly a constant with respect to the temperature in both the standard NJL model and the lattice-improved NJL model. In the interval of $100\,\text{MeV}\lesssim T \lesssim 200\,\text{MeV}$, the value of $\Delta\overline{M}_{\sigma,\pi^0}^{\ssp{\parallel}}$ or $\Delta\overline{M}_{\sigma,\pi^0}^{\ssp{\perp}}$ is sharply depressed by the growth of the temperature, which means that the chiral symmetry is restoring in this temperature range. 
At very high temperature ($T\gtrsim200$ MeV), the screening masses of $\pi^0$ and $\sigma$ mesons become degenerate with
a good approximation, i.e.
$\Delta\overline{M}_{\sigma,\pi^0}^{\ssp{\parallel}}=0$ or $\Delta\overline{M}_{\sigma,\pi^0}^{\ssp{\perp}}=0$, and this fact signifies the restoration of chiral
symmetry accordingly. Therefore, we can resort to the meson screening mass difference $\Delta\overline{M}_{\sigma,\pi^0}^{\ssp{\parallel}}$ (or $\Delta\overline{M}_{\sigma,\pi^0}^{\ssp{\perp}}$) as an order parameter instead of the quark condensate to define
the critical temperature of the chiral phase transition.


\begin{figure*}[htbp]
	\includegraphics[scale=0.25]{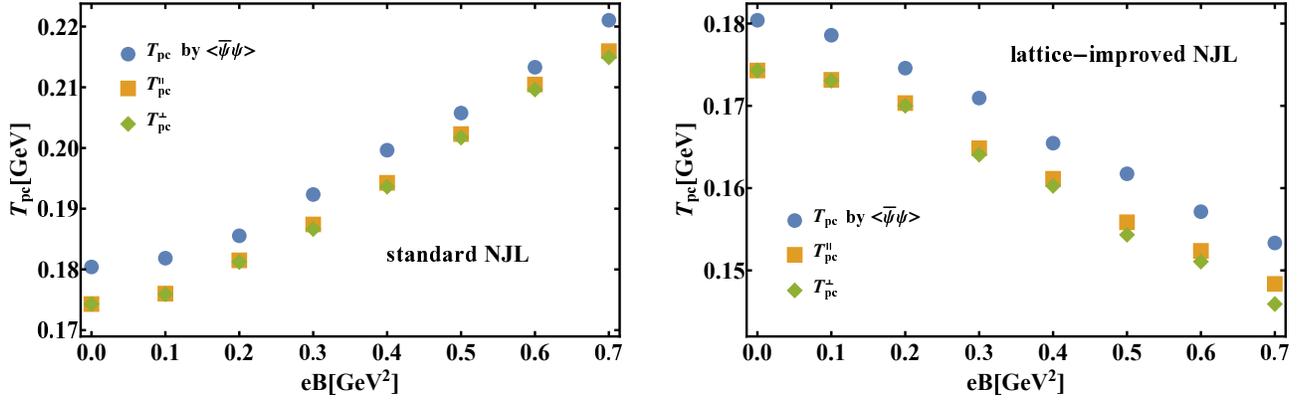}
	\caption{\label{fig9}The pseudo-critical temperatures $T_{pc}^{\parallel}$ and $T_{pc}^{\perp}$,
		as well as $T_{pc}$, as functions of $eB$ in the standard (left panel) and the lattice-improved (right panel) NJL models.}
\end{figure*}
\begin{figure*}[htbp]
	\includegraphics[scale=0.25]{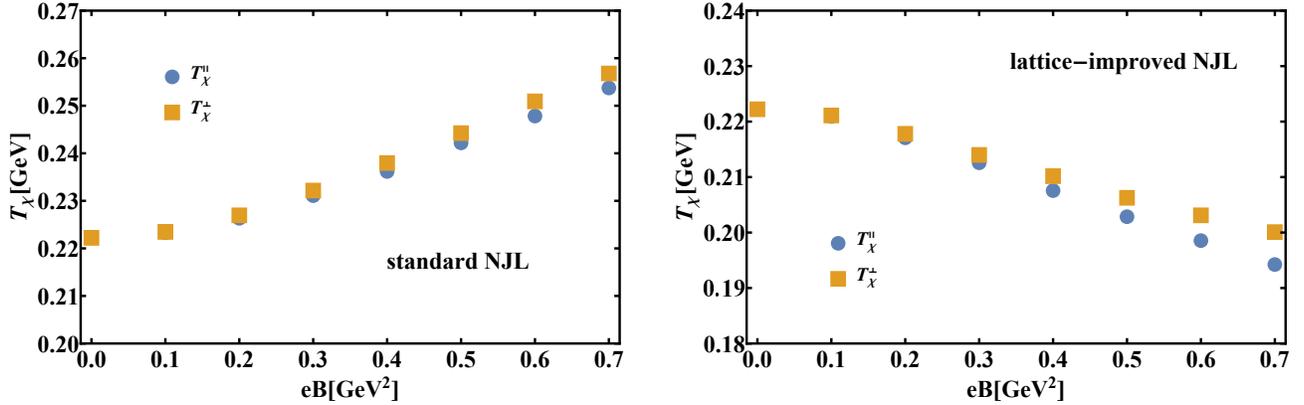}
	\caption{\label{fig10}The critical temperatures $T_{\chi}^{\parallel}$ and $T_{\chi}^{\perp}$,
		as functions of $eB$ in the standard (left panel) and the lattice-improved (right panel) NJL models.}
\end{figure*}

\begin{figure*}[htbp]
	\includegraphics[scale=0.25]{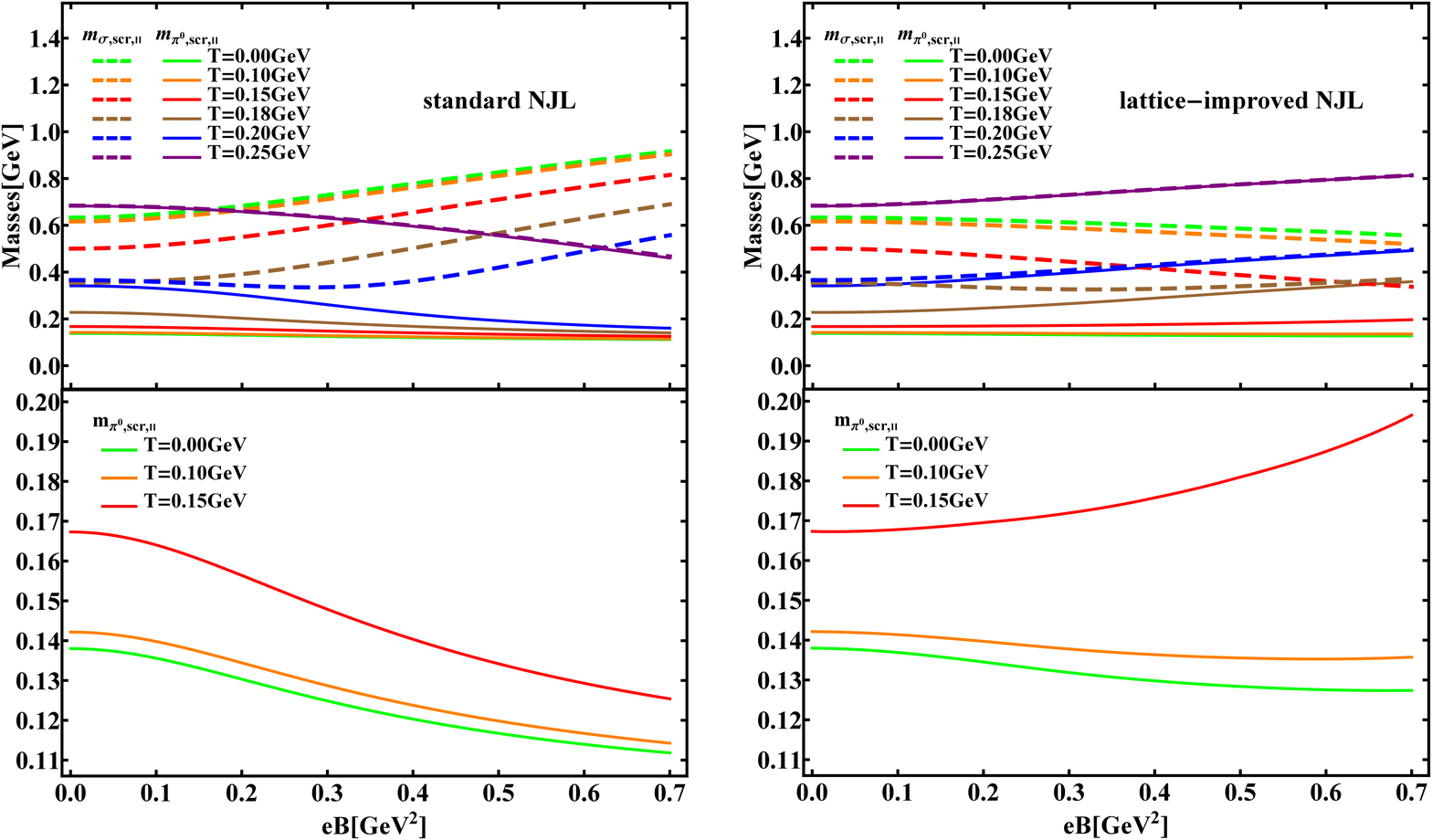}
	\caption{\label{fig11}Left panel: The longitudinal screening masses of $\pi^0$ and $\sigma$ mesons as functions of $eB$ at fixed $T=0.00$, $0.10$, $0.15$, $0.18$, $0.20$ and $0.25$ $\rm{GeV}$ in the standard NJL model, and the curves for neutral pion at low temperatures are plotted on the bottom for visibility. Right panel: The same quantities in the lattice-improved NJL model.}
\end{figure*}

Note that we define two critical temperatures by using the $\pi^0-\sigma$ screening mass difference: one is the inflection point of $\Delta\overline{M}_{\sigma,\pi^0}^{\ssp{\parallel}}$ (or $\Delta\overline{M}_{\sigma,\pi^0}^{\ssp{\perp}}$) curve, which is identified with the pseudo-critical temperature $T_{pc}^{\parallel}$ (or $T_{pc}^{\perp}$)
of the chiral crossover, in analogy to the prescription for the quark condensate, namely
\beq\label{Eq41}
\left.\dfrac{\partial^2\Delta\overline{M}_{\sigma,\pi^0}^{\ssp{\parallel}}}{\partial T^2}\right|_{T=T_{pc}^{\parallel}}=0~\text{or}~ \left.\dfrac{\partial^2\Delta\overline{M}_{\sigma,\pi^0}^{\ssp{\perp}}}{\partial T^2}\right|_{T=T_{pc}^{\perp}}=0~;
\eeq
the other is the merging point of $m_{\ssp{\pi^0},scr,\ssp{\parallel}}$ (or $m_{\ssp{\pi^0},scr,\ssp{\perp}}$) and $m_{\ssp{\sigma},scr,\ssp{\parallel}}$ (or $m_{\ssp{\sigma},scr,\ssp{\perp}}$), which is the critical temperature $T_{\chi}^{\parallel}$ (or $T_{\chi}^{\perp}$) signifying the chiral symmetry restoration and the definition reads
\beq\label{Eq42}
\left.\Delta\overline{M}_{\sigma,\pi^0}^{\ssp{\parallel}}\right|_{T=T_{\chi}^{\parallel}}=0~\text{or}~ \left.\Delta\overline{M}_{\sigma,\pi^0}^{\ssp{\perp}}\right|_{T=T_{\chi}^{\perp}}=0~.
\eeq
Nevertheless, since the chiral phase transition is actually a crossover owing to non-zero current quark mass, a small arbitrary value $\epsilon$ can be chosen to determine the critical temperatures $T_{\chi}^{\parallel}$ and $T_{\chi}^{\perp}$, e.g., $\left.\Delta\overline{M}_{\sigma,\pi^0}^{\ssp{\parallel}}\right|_{T=T_{\chi}^{\parallel}}=\left.\Delta\overline{M}_{\sigma,\pi^0}^{\ssp{\perp}}\right|_{T=T_{\chi}^{\perp}}=\epsilon=10^{-2}$ in our numerical calculations.
With the help of $T_{pc}^{\parallel}$ ($T_{pc}^{\perp}$) and $T_{\chi}^{\parallel}$ ($T_{\chi}^{\perp}$), the effects of the external magnetic field on these critical temperatures can be analyzed in the following.

On the one hand, the pseudo-critical temperatures $T_{pc}^{\parallel}$ and $T_{pc}^{\perp}$ as functions of the magnetic field are shown in Fig.~\ref{fig9}. And the pseudo-critical temperature $T_{pc}$ defined by the inflection points of the quark condensate from Fig.~\ref{fig3} is also sketched for comparison. In general, the $eB$ dependences of $T_{pc}^{\parallel}$ and $T_{pc}^{\perp}$ are essentially coincident with that of $T_{pc}$ not only in the standard NJL model but also in the lattice-improved NJL model, although somehow less than $T_{pc}$. It means that the normalized screening mass difference $\Delta\overline{M}_{\sigma,\pi^0}^{\ssp{\parallel}}$ and $\Delta\overline{M}_{\sigma,\pi^0}^{\ssp{\perp}}$ can effectively describe the chiral phase transition also. Evidently, in the left panel of Fig.~\ref{fig9}, the value of $T_{pc}^{\parallel}$ ($T_{pc}^{\perp}$) becomes larger and larger with the increase of the magnetic field, which is in  agreement with the result obtained by the quark condensate in the standard NJL model. Whereas, in the right panel of Fig.~\ref{fig9}, the decrease of $T_{pc}^{\parallel}$ ($T_{pc}^{\perp}$) with the increasing $eB$ reflects the IMC from the LQCD simulations.

On the other hand, the $eB$ dependences of $T_{\chi}^{\parallel}$ and $T_{\chi}^{\perp}$ are sketched in Fig.~\ref{fig10}. The behaviors of them are quite similar to those of the pseudo-critical temperatures in Fig.~\ref{fig9}, although the values of $T_{\chi}^{\parallel}$ ($T_{\chi}^{\perp}$) are obviously larger than $T_{pc}^{\parallel}$ ($T_{pc}^{\perp}$) in both the standard NJL model and the lattice-improved NJL model. In the left panel of Fig.~\ref{fig10},
the increase of $T_{\chi}^{\parallel}$ and $T_{\chi}^{\perp}$ with the growing $eB$ reflects the MC in the standard NJL model. And in the right panel of Fig.~\ref{fig10}, their opposite behaviors varying with $eB$ indicate the IMC effect, which can be well reproduced in the lattice-improved NJL model. Additionally, $T_{pc}^{\parallel}$ is almost the same as $T_{pc}^{\perp}$ at any value of $eB$ as shown in Fig.~\ref{fig9}, since they have nothing to do with the anisotropy of spacetime caused by the external magnetic field. Whereas, in Fig.~\ref{fig10}, the difference between $T_{\chi}^{\parallel}$ and $T_{\chi}^{\perp}$ becomes larger and larger as $eB$ increases, and $T_{\chi}^{\perp}$ is always not less than  $T_{\chi}^{\parallel}$. This is because of the definitions of $T_{\chi}^{\parallel}$ and $T_{\chi}^{\perp}$ in the latter case: it is not difficult to find that in the low-momentum expansion, 
 \beq\label{Eq36}
m^2_{\ssp{\sigma},scr,\ssp{\parallel}}-m^2_{\ssp{\pi^0},scr,\ssp{\parallel}}=4m^2
\eeq
and
\beq\label{Eq37}
m_{\ssp{\sigma},scr,\ssp{\perp}}^2-m_{\ssp{\pi^0},scr,\ssp{\perp}}^2=4m^2\times
\dfrac{\mathrm{I}_{2,\ssp{\parallel}}(0)}{\mathrm{I}_{2,\ssp{\perp}}(0)}
\eeq
where $\mathrm{I}_{2,\ssp{\parallel}}(0)=\lim \limits_{q \to 0} \mathrm{I}_{2,\ssp{\parallel}}(q)$ and
$\mathrm{I}_{2,\ssp{\perp}}(0)=\lim \limits_{q \to 0} \mathrm{I}_{2,\ssp{\perp}}(q)$, which are good
approximations in the low temperature region~\cite{Sheng:2020hge}. In this ansatz, we could have $\dfrac{(m_{\ssp{\sigma},scr,\ssp{\perp}}^2-m_{\ssp{\pi^0},scr,\ssp{\perp}}^2)}{(m_{\ssp{\sigma},scr,\ssp{\parallel}}^2-m_{\ssp{\pi^0},scr,\ssp{\parallel}}^2)}=\dfrac{\mathrm{I}_{2,\ssp{\parallel}}(0)}{\mathrm{I}_{2,\ssp{\perp}}(0)}\geqslant 1$ and thus $\dfrac{T_{\chi}^{\perp}}{T_{\chi}^{\parallel}}\geqslant 1$, both of which grow with the increase of $eB$.

\subsubsection{Results at fixed $T$}\label{sec:four,three,two}

\begin{figure*}[htbp]
	\includegraphics[scale=0.25]{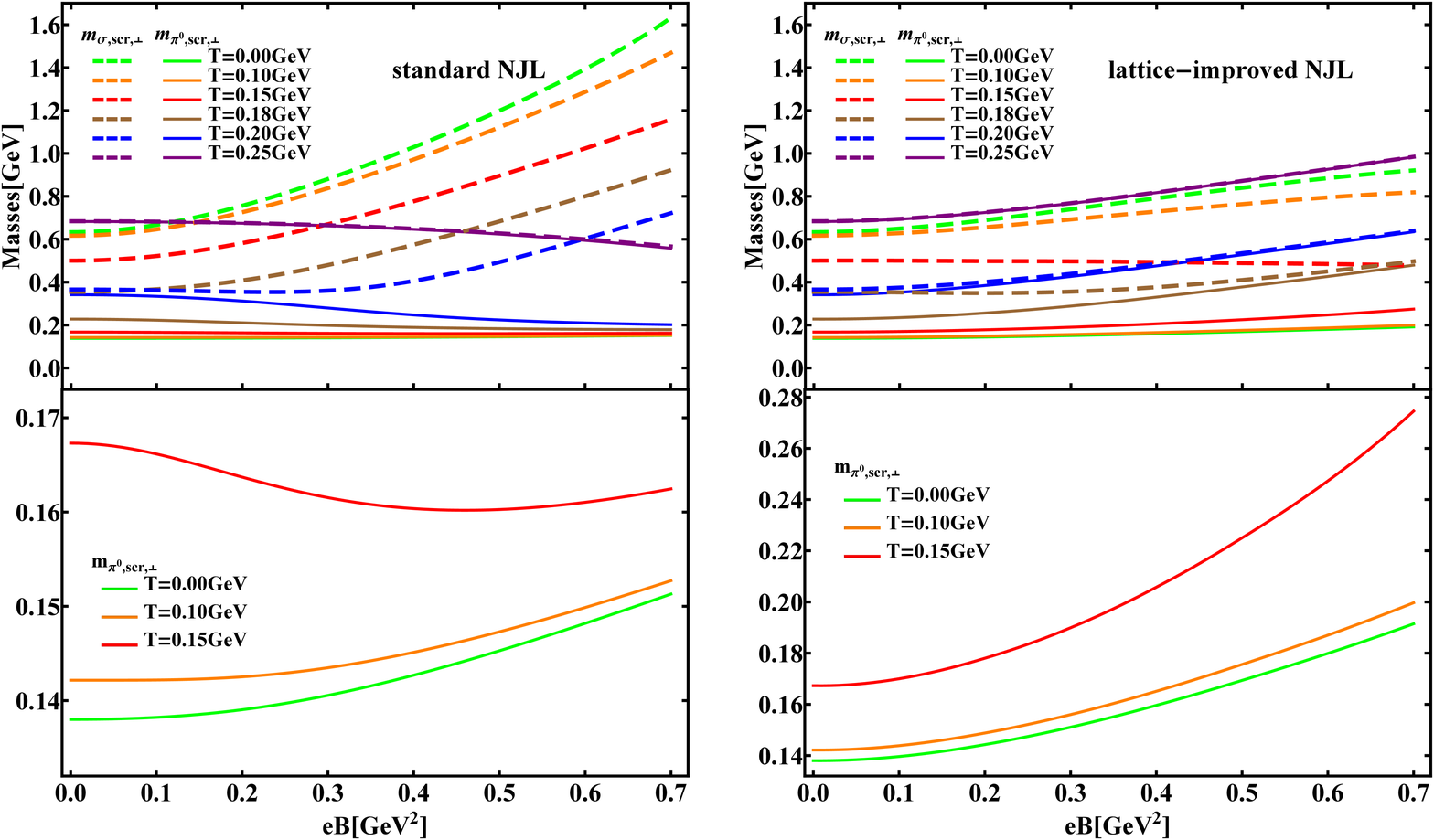}
	\caption{\label{fig12}Left panel: The transverse screening masses of $\pi^0$ and $\sigma$ mesons as functions of $eB$ at fixed $T=0.00$, $0.10$, $0.15$, $0.18$, $0.20$ and $0.25$ $\rm{GeV}$ in the standard NJL model, and the curves for neutral pion at low temperatures are plotted on the bottom for visibility. Right panel: The same quantities in the lattice-improved NJL model.}
\end{figure*}

In this subsection, we first continue to present the $eB$ dependences of $m_{\pi^0, scr, \parallel}$ and $m_{\sigma, scr, \parallel}$ at fixed $T=0.00$, $0.10$, $0.15$, $0.18$, $0.20$ and $0.25$ $\rm{GeV}$ in Fig.~\ref{fig11}, in order to illustrate the effects of the IMC on the meson screening masses further. Specifically, in the standard NJL model, $m_{\pi^0,scr,\ssp{\parallel}}$ reduces monotonously with the growth of $eB$ at either low or high temperatures, while in the lattice-improved NJL model, $m_{\pi^0,scr,\ssp{\parallel}}$ still decreases with $eB$ at low temperatures (e.g., at $T=0\,\text{MeV}$ and $T=100\,\text{MeV}$), despite that the extents of the decrease are smaller than those in the standard NJL model, but when the temperature is high enough (e.g., $T\geq150\,\text{MeV}$), $m_{\pi^0,scr,\ssp{\parallel}}$ is converted to increase with $eB$. 

As concerning $m_{\sigma,scr,\ssp{\parallel}}$, it is interesting to find that $m_{\sigma,scr,\ssp{\parallel}}$ increases in the standard NJL model but decreases in the lattice-improved NJL model at low temperatures, as $eB$ grows. This is because that, according to Eq.~(\ref{Eq36}), since $m_{\ssp{\pi^0},scr,\ssp{\parallel}}$ is almost a constant with respect to $eB$ at $T\lesssim150$ $\text{MeV}$, the behavior of $m_{\sigma,scr,\ssp{\parallel}}$ is mainly determined by that of the constituent quark mass $m$, and $m$ in the lattice-improved NJL model decreases with $eB$ as a result of the decreasing behavior of $G(eB)$ with regard to $eB$, in contrary to $m$ in the standard NJL model. At very high temperatures, e.g. $T=250\text{MeV}$, $m_{\sigma,scr,\ssp{\parallel}}$ obviously becomes degenerate with $m_{\pi^0,scr,\ssp{\parallel}}$, and decreases in the standard NJL model but increases in the lattice-improved NJL model with the increase of $eB$ instead. When around the chiral crossover, we find that $m_{\ssp{\sigma},scr,\ssp{\parallel}}$ increases with $eB$ at $T=180$ $\text{MeV}$, and first slightly decreases and then increases with $eB$ at $T=200$ $\text{MeV}$ in the standard NJL model, as shown in the left panel of Fig.~\ref{fig11}. However, as depicted by the right panel of Fig.~\ref{fig11}, $m_{\sigma,scr,\ssp{\parallel}}$ first slightly decreases and then increases with $eB$ at $T=180$ $\text{MeV}$, and turns to increase with $eB$ monotonously at $T=200$ $\text{MeV}$ in the lattice-improved NJL model. 

Next, the plots of $m_{\pi^0, scr, \ssp{\perp}}$ and $m_{\sigma, scr, \ssp{\perp}}$ as functions of $eB$ at fixed temperatures are shown in Fig.~\ref{fig12}. One interesting thing is that at low temperatures (i.e., $T=0$ and $100$ $\text{MeV}$), $m_{\ssp{\pi^0},scr,\ssp{\perp}}$ in either the standard NJL model or the lattice-improved NJL model is enhanced by the magnetic field, in contrast to
$m_{\pi^0, scr, \ssp{\parallel}}$. But the rising of temperature makes the situation different: in the standard NJL model, $m_{\ssp{\pi^0},scr,\ssp{\perp}}$ turns to first decrease and then increase with
the increasing of $eB$ at $T=150$ $\text{MeV}$, and eventually becomes reduced with $eB$ at high temperatures (e.g., $T=180$, $200$ and $250$ $\text{MeV}$); while in the lattice-improved NJL model, $m_{\ssp{\pi^0},scr,\ssp{\perp}}$ remains increasing with $eB$ as the temperature grows. As for $m_{\ssp{\sigma},scr,\ssp{\perp}}$, when in the standard NJL model, the $eB$ dependences of $m_{\ssp{\sigma},scr,\ssp{\perp}}$ at different temperatures are qualitatively similar to those of $m_{\ssp{\sigma},scr,\ssp{\parallel}}$. On the other hand, when in the lattice-improved NJL model, although $m_{\ssp{\sigma},scr,\ssp{\perp}}$ and $m_{\ssp{\sigma},scr,\ssp{\parallel}}$ behave similarly to each other as $eB$ increases at enough high temperatures ($T\gtrsim150$ $\text{MeV}$), there are some difference between them at low temperatures (e.g. $T=0$ and $100$ $\text{MeV}$): $m_{\ssp{\sigma},scr,\ssp{\perp}}$ increases with $eB$, while $m_{\ssp{\sigma},scr,\ssp{\parallel}}$ hold decreasing with $eB$.

\begin{figure}[htbp]
	\includegraphics[scale=0.25]{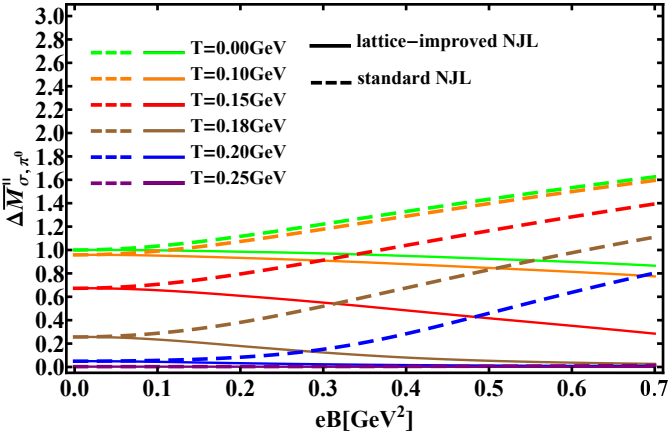}
	\caption{\label{fig13}The normalized longitudinal screening mass difference  $\Delta\overline{M}_{\sigma,\pi^0}^{\ssp{\parallel}}$ as a function of $eB$ at fixed $T=0.0$, $0.10$, $0.15$, $0.18$, $0.20$ and $0.25$ $\rm{GeV}$.}
\end{figure}
\begin{figure}[htbp]
	\includegraphics[scale=0.25]{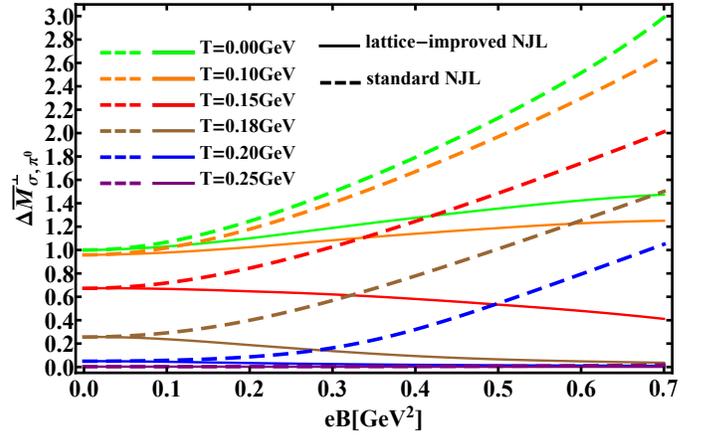}
	\caption{\label{fig14}The normalized transverse screening mass difference  $\Delta\overline{M}_{\sigma,\pi^0}^{\ssp{\perp}}$ as a function of $eB$ at fixed $T=0.0$, $0.10$, $0.15$, $0.18$, $0.20$ and $0.25$ $\rm{GeV}$.}
\end{figure}

And then, by making use of the numerical results in Figs.~\ref{fig11} and \ref{fig12}, we examine the $eB$ dependences of $\Delta\overline{M}_{\sigma,\pi^0}^{\ssp{\parallel}}$ and $\Delta\overline{M}_{\sigma,\pi^0}^{\ssp{\perp}}$ at different temperatures in Fig.~\ref{fig13} and Fig.~\ref{fig14}, respectively. Obviously, in the standard NJL model, only magnetic catalysis can be uncovered by $\Delta\overline{M}_{\sigma,\pi^0}^{\ssp{\parallel}}$ and $\Delta\overline{M}_{\sigma,\pi^0}^{\ssp{\perp}}$ that are enhanced by the increase of $eB$ at arbitrary temperature. However, both of them decrease with the increasing $eB$ at high temperatures (below $T_{\chi}^{\parallel}$ or $T_{\chi}^{\perp}$, of course) in the lattice-improved NJL model, which is consistent with the corresponding inverse magnetic catalysis acquired by the chiral condensate. Moreover, at low temperatures, it is not difficult to find that in both the standard and lattice-improved NJL models, $\Delta\overline{M}_{\sigma,\pi^0}^{\ssp{\parallel}}$ shows same $eB$ dependence as $m$ according to Eq.~(\ref{Eq36}), while $\Delta\overline{M}_{\sigma,\pi^0}^{\ssp{\perp}}$ acts much like the quark condensate as $eB$ grows.

\begin{figure*}[htbp]
	\includegraphics[scale=0.25]{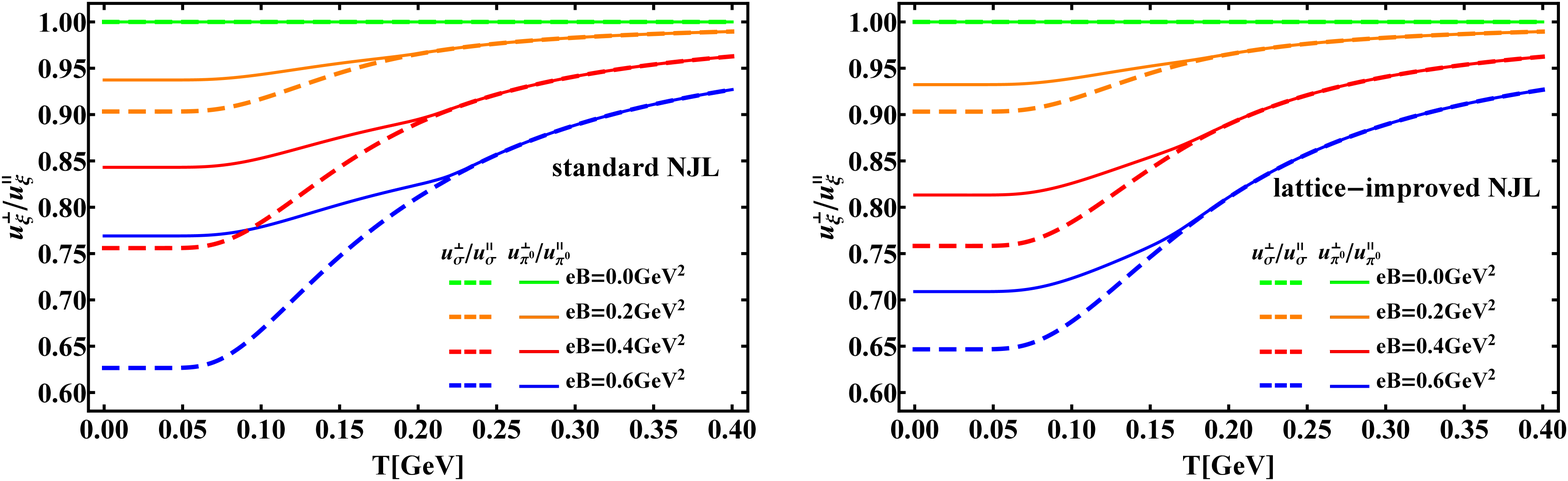}
	\caption{\label{fig15}$u_{\xi}^{\ssp{\perp}}/u_{\xi}^{\ssp{\parallel}}$ as a function of $T$ at fixed $eB=0.0$, $0.2$, $0.4$ and $0.6$ $\rm{GeV^2}$ in the standard NJL model (left panel) and the lattice-improved NJL model (right panel), respectively.}
\end{figure*}
\begin{figure*}[htbp]
	\includegraphics[scale=0.25]{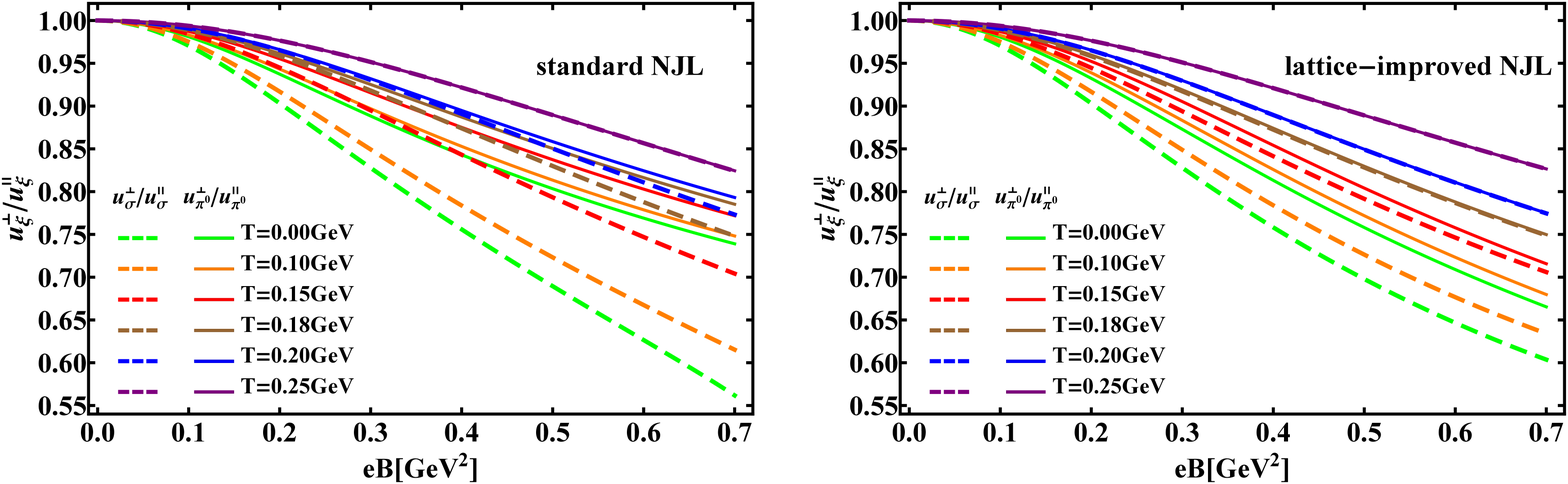}
	\caption{\label{fig16}$u_{\xi}^{\ssp{\perp}}/u_{\xi}^{\ssp{\parallel}}$ as a function of $eB$ at fixed $T=0.0$, $0.10$, $0.15$, $0.18$, $0.20$ and $0.25$ $\rm{GeV}$ in the standard NJL model (left panel) and the lattice-improved NJL model (right panel), respectively.}
\end{figure*}

Furthermore, by comparing Fig.~\ref{fig11} with Fig.~\ref{fig12}, we can find there are quantitative differences between $m_{\pi^0, scr, \ssp{\parallel}}$ ($m_{\sigma, scr, \ssp{\parallel}}$) and $m_{\pi^0, scr, \ssp{\perp}}$ ($m_{\sigma, scr, \ssp{\perp}}$) at fixed temperature in the same model, which are related to the anisotropy of space-time caused by the background magnetic field and will be discussed in the following subsection.

\subsection{\label{sec:four,four}The ratio of sound velocities $u_{\xi}^{\ssp{\perp}}/u_{\xi}^{\ssp{\parallel}}$}

As mentioned in Ref.~\cite{Sheng:2020hge}, the degree of the asymmetry between the longitudinal and transverse directions, stemming from the external magnetic field, can be measured by the ratio of sound velocities $u_{\xi}^{\ssp{\perp}}/u_{\xi}^{\ssp{\parallel}}=m_{\ssp{\xi},scr,\ssp{\parallel}}/m_{\ssp{\xi},scr,\ssp{\perp}}$, with the definitions
$u_{\xi}^{\ssp{\parallel}}=\dfrac{m_{\ssp{\xi},pole}}{m_{\ssp{\xi},scr,\ssp{\parallel}}}$ and $u_{\xi}^{\ssp{\perp}}=\dfrac{m_{\ssp{\xi},pole}}{m_{\ssp{\xi},scr,\ssp{\perp}}}$. Thus, Fig.~\ref{fig15} and Fig.~\ref{fig16} show the $T$ and $eB$ dependences of the ratio $u_{\xi}^{\ssp{\perp}}/u_{\xi}^{\ssp{\parallel}}$ for $\xi=\pi^0$ and $\sigma$ mesons, respectively. According to our arguments about the Lorentz symmetry breaking and the causality in Ref.~\cite{Sheng:2020hge}, it is expected that $u_{\xi}^{\ssp{\perp}}/u_{\xi}^{\ssp{\parallel}}=1$ at $eB=0$ and $u_{\xi}^{\ssp{\perp}}/u_{\xi}^{\ssp{\parallel}}<1$ at $eB\neq0$ at zero and finite temperature, which are in agreement with the results in Fig.~\ref{fig15} and Fig.~\ref{fig16}. Furthermore, $u_{\pi^0}^{\ssp{\perp}}/u_{\pi^0}^{\ssp{\parallel}}$ and $u_{\sigma}^{\ssp{\perp}}/u_{\sigma}^{\ssp{\parallel}}$ are both reduced by the increasing $eB$ because of the enhancement of the symmetry breaking in coordinate space by the magnetic field, no matter in the standard NJL model or in the lattice-improved NJL model. Note that the $eB$ dependence of $u_{\pi^0}^{\ssp{\perp}}/u_{\pi^0}^{\ssp{\parallel}}$ in the standard NJL model as shown in the left panel of Fig.~\ref{fig16} is qualitatively in accordance with the results in Ref.~\cite{Sheng:2020hge}.

In more details, it is shown in Fig.~\ref{fig15} that, at fixed $eB=0.2$, $0.4$ and $0.6$ $\rm{GeV^2}$, both $u_{\pi^0}^{\ssp{\perp}}/u_{\pi^0}^{\ssp{\parallel}}$ and $u_{\sigma}^{\ssp{\perp}}/u_{\sigma}^{\ssp{\parallel}}$ are temperature independent when $T \lesssim 50$ $\text{MeV}$, and then gradually increase to unity as the temperature grows when $T\gtrsim 50$ $\text{MeV}$. Obviously, it agrees with the result in Ref.~\cite{Sheng:2020hge} that, $u_{\xi}^{\ssp{\perp}}/u_{\xi}^{\ssp{\parallel}}$ depends on the magnetic field strength only at low temperatures, where the screening effect of the temperature can be decoupled from that of the magnetic field, but when the temperature becomes sufficiently high, the space symmetry broken by the magnetic field will be recovered by the random thermal motion with the increasing of $T$~\cite{Wang:2017vtn,Sheng:2020hge}. On the other side, as depicted in Fig.~\ref{fig16}, it is found that at low temperatures there is always some discrepancy between $u_{\pi^0}^{\ssp{\perp}}/u_{\pi^0}^{\ssp{\parallel}}$ and $u_{\sigma}^{\ssp{\perp}}/u_{\sigma}^{\ssp{\parallel}}$ (i.e., $u_{\sigma}^{\ssp{\perp}}/u_{\sigma}^{\ssp{\parallel}}<u_{\pi^0}^{\ssp{\perp}}/u_{\pi^0}^{\ssp{\parallel}}<1$) at finite $eB$ on account of their mass difference, and when $T \gtrsim T_{pc}^{\parallel}$ or $T_{pc}^{\perp}$, $u_{\pi^0}^{\ssp{\perp}}/u_{\pi^0}^{\ssp{\parallel}}$ and $u_{\sigma}^{\ssp{\perp}}/u_{\sigma}^{\ssp{\parallel}}$ become degenerate along with the restoration of chiral symmetry. 

\begin{figure}[htbp]
	\includegraphics[scale=0.25]{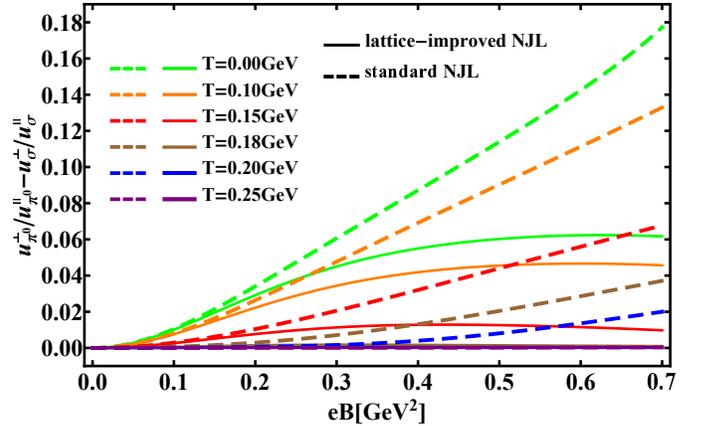}
	\caption{\label{fig17}The ratio difference $u_{\pi^0}^{\ssp{\perp}}/u_{\pi^0}^{\ssp{\parallel}}- u_{\sigma}^{\ssp{\perp}}/u_{\sigma}^{\ssp{\parallel}}$ as a function of $eB$ at fixed $T=0.0$, $0.10$, $0.15$, $0.18$, $0.20$ and $0.25$ $\rm{GeV}$ in the standard NJL model and the lattice-improved NJL model.}
\end{figure}

Finally, we study the $eB$ dependence of the ratio difference $u_{\pi^0}^{\ssp{\perp}}/u_{\pi^0}^{\ssp{\parallel}}- u_{\sigma}^{\ssp{\perp}}/u_{\sigma}^{\ssp{\parallel}}$, which is shown in Fig.~\ref{fig17}. And we could find that in the low-temperature region (e.g.,$T =0$, $100$ and $150$ $\text{MeV}$), when $eB \gtrsim 0.5$ $\text{GeV}^2$, the ratio difference $u_{\pi^0}^{\ssp{\perp}}/u_{\pi^0}^{\ssp{\parallel}}- u_{\sigma}^{\ssp{\perp}}/u_{\sigma}^{\ssp{\parallel}}$ in the standard NJL model continues increasing with $eB$, while the one in the lattice-improved NJL model reaches a saturation,
although both of them are enlarged with the increase of the magnetic field when $eB < 0.5$ $\text{GeV}^2$. Additionally, at a certain
$eB$, $u_{\pi^0}^{\ssp{\perp}}/u_{\pi^0}^{\ssp{\parallel}}- u_{\sigma}^{\ssp{\perp}}/u_{\sigma}^{\ssp{\parallel}}$ in the lattice-improved NJL model is smaller than that in the standard NJL model at low temperatures, which is mainly attributed to the decreasing behavior of $G(eB)$ with respect to $eB$ in the lattice-improved NJL model.

\section{\label{sec:five}Summary and Conclusions}

In this paper, we incorporate IMC effectively in the lattice-improved two-flavor NJL model by introducing an eB-dependent coupling constant $G(eB)$ to the four-quark interaction~\cite{Endrodi:2019whh}. The eB-dependence of $G(eB)$ is determined by utilizing the magnetic field dependent constituent quark masses inferred from the magnetized baryon mass spectrum which is evaluated in LQCD. The lattice-improved NJL model is shown to exhibit IMC at high temperatures and a reduction of the pseudo-critical temperature as the magnetic field grows, which are consistent with the lattice results in Refs.~\cite{Bali:2011qj,Bali:2012zg}.

In order to investigate the effects of IMC on the meson screening masses, we analyze the longitudinal and transverse screening masses of neutral pion and sigma meson in terms of the lattice-improved NJL model. For comparison, we also calculate meson screening masses in the standard NJL model that the coupling constant is fixed at $G(eB=0)$. For the sake of the decreasing behavior of the coupling constant $G(eB)$ with increasing $eB$, the monotonicity of $m_{\ssp{\pi^0},scr,\ssp{\parallel}}$ and $m_{\ssp{\pi^0},scr,\ssp{\perp}}$ for neutral pions with respect to the magnetic field in the lattice-improved NJL model is different from that in the standard NJL model when the temperature is adequately high, i.e., $T\gtrsim 150\text{MeV}$. As concerning sigma mesons, because of the same reason above, the monotonicity of $m_{\ssp{\sigma},scr,\ssp{\parallel}}$ with regard to $eB$ in the lattice-improved NJL model also differs from that in the standard NJL model at low temperatures. Meanwhile, although $m_{\ssp{\sigma},scr,\ssp{\perp}}$ in both the lattice-improved NJL model and the standard NJL model monotonously increases as $eB$ grows in the low temperature regime, the extent of increase in the lattice-improved NJL model is smaller than that in the standard NJL model. 

In particular, it is interesting to find the fact that, when around transition temperature, the $\sigma-\pi^0$ meson screening mass differences $\Delta{m}_{\sigma,\pi^0}^{\ssp{\parallel}}$ and $\Delta{m}_{\sigma,\pi^0}^{\ssp{\perp}}$ increase with
$eB$ in the standard NJL model but decrease with $eB$ in the lattice-improved NJL model, which is in 
consistent with the scenario predicted by the quark condensate. By the aid of these screening mass differences, $T_{pc}^{\parallel}$ and $T_{pc}^{\perp}$, as well as $T_{\chi}^{\parallel}$ and $T_{\chi}^{\perp}$ are defined as (pseudo)critical temperatures of chiral transition. And we find that in both the standard and the lattice-improved NJL models, the eB-dependences of $T_{pc}^{\parallel}$ and $T_{pc}^{\perp}$ as well as $T_{\chi}^{\parallel}$ and $T_{\chi}^{\perp}$, are in good agreement with that of $T_{pc}$ by the quark condensate. Hence, it implies that exploring behaviors of the screening mass differences of chiral partners (e.g. $\pi^0$ and $\sigma$ mesons) may help to uncover the underlying mechanism of IMC. And it is
expected that our relevant predictions above could be proved or disproved by lattice QCD simulations in the near future, so that we can examine validity and reliability of the hypothesis that the magnetic field dependent coupling constant in the NJL model gives rise to IMC in a phenomenological manner.

\acknowledgements
The authors thank Danning Li, Song He, Yong-Liang Ma and Shinya Matsuzaki for useful discussion. L.Y. acknowledges the kind hospitality of the Department of physics and Siyuan Laboratory at Jinan University. The work of L.Y. is supported by the
NSFC under Grant No. 11605072 and 12047569, and the Seeds Funding of Jilin University.
X.W. is supported by
the start-up funding No. 4111190010 of Jiangsu University.

\bibliography{bib}
\end{document}